\documentclass[11pt]{article}

\usepackage{epsf,latexsym,amsmath,amssymb,amsfonts,citesort,graphicx,lscape,amscd,epsfig,psfrag,subfigure,dsfont,cite,array,dsfont}

\newcommand{\QED}{\hfill $\blacksquare$}

\newcommand{\p}{P_{av}}
\newcommand{\pa}{\mbox{$P_{av}$}}
\newcommand{\ou}{\Pi(\p,R(\p))}

\newtheorem{prop}{Proposition}

\newtheorem{thm}[prop]{Theorem}
\newtheorem{lemma}[prop]{Lemma}
\newcommand{\s}{{\mathsf{SNR}}}

\title{On the Asymptotic Performance of Multiple Antenna Channels with Fast Channel Feedback}
\author{Ahmad Khoshnevis and Ashutosh Sabharwal}
\date{}

\begin{document}

\maketitle

\begin{abstract}
In this paper, we analyze the asymptotic performance of multiple antenna channels where the transmitter has either perfect or
finite bit channel state information. Using the diversity-multiplexing tradeoff to characterize the system performance, we
demonstrate that channel feedback can fundamentally change the system behavior. Even one-bit of information can increase the
diversity order of the system compared to the system with no transmitter information. In addition, as the amount of channel
information at the transmitter increases, the diversity order for each multiplexing gain increases and goes to infinity for
perfect transmitter information. The major reason for diversity order gain is a ``location-dependent" temporal power control,
which adapts the power control strategy based on the average channel conditions of the channel.
\end{abstract}

\section{Introduction}

 The importance of channel state information at transmitter has been extensively studied, see for
 example~\cite{Tel95,GV97,CTB99,NLTW98,MSEA03,LHS03,Raj02}. In fading channels, channel state information at the transmitter
 leads to significant gains in outage probability over systems with no information at the transmitter. There are many widely
 used methods to exploit the transmitter channel information like power control and beamforming. However, the comparative
 performance  of these different methods is still unclear. In this paper, we characterize the diversity order in systems with
 perfect or imperfect channel state information at the transmitter, with the aim of identifying techniques which yield maximum
 benefit in outage performance.


We study systems in which the transmitter has causal channel information and adapts its actions based on the current channel
conditions. Our main contributions are four-fold. First, we study the performance with \emph{perfect} channel information at the
transmitter and receiver, and show that the diversity order of optimal power control is infinite at all multiplexing gains, for
many antenna configurations. To enable fair comparisons with systems which have no or imperfect channel information at the
transmitter, the outage definition is generalized to include instances in which no information is transmitted by the sender.
Thus, the outage event contains all the cases in which either the transmitter sends no information or the information is sent
but corrupted by the channel.

Second, we refine our analysis further by decoupling the optimal power control into temporal power control and spatial power
control. In temporal power control, all eigenvectors of the channel receive the same equal power, with total power depending on
the current channel conditions. And in spatial power control, a short term power constraint is employed leading to same total
power for each channel condition but adaptive power allocation along different eigenvectors of the channel. We show that most of
the gain of optimal power control stems from temporal power control. In fact, spatial power control has the same
diversity-multiplexing tradeoff as the system with no channel information at the transmitter. To achieve the gain from temporal
power control, an adaptive power control is critical which adapts based on average condition of forward channel.

Third, we derive lower bounds on diversity order for systems where only finite number of feedback bits about the channel are
made available to the transmitter. The analysis is limited to only temporal power control since only temporal power control
provides increase in diversity (as shown by the above mentioned perfect channel information analysis). To aid closed form
analysis, we develop a simple finite bit quantizer which can be computed in a recursive manner. We show that finite bit channel
information leads to a substantial (though finite) increase in diversity order for each multiplexing gain, when compared to
system with no channel information~\cite{ZT03}. And as the number of feedback bits increase, the diversity order increases
unboundedly for all but the maximum multiplexing gain, and approaches the perfect information case. The proof of the finite rate
feedback relies on approximations of order statistics of eigenvalues of Gaussian matrices, which are of independent interest.

Fourth and last, we also consider joint rate and power control with finite number of feedback bits, and show that a non-zero
diversity gain is possible for highest multiplexing gain even with imperfect transmitter information. This contrasts the
temporal power control methods with finite rate feedback which have zero diversity order for full multiplexing gain.


Several recent works~\cite{OZF05,HG05} have derived the diversity-multiplexing tradeoff with channel information at the
transmitter, for beamforming like spatial resource allocation methods. We note that the work in~\cite{GCD05} is closest to our
work in spirit with one major difference. In~\cite{GCD05}, authors consider the impact of using $L$-rounds of ARQ in a MIMO
channel, and show that additional delay changes the diversity-multiplexing behaviour in much the same way as our results. Each
ARQ round provides one-bit information about the channel output and hence $L$-rounds provide $L$ bits of information. In
contrast, we consider impact of channel information, not the channel output in this paper. In our opinion, the two works are
complementary and could form the basis of answering a more fundamental question. If we can send only $b$-bits of feedback per
unit time, then how should those $b$ bits be selected. The obvious options include quantizing the channel and channel output
(equivalently decoder state). The difference between the two possibilities is that channel feedback occurs before the data
transmission and channel output feedback after the data transmission. A priori it is unclear which form of feedback provides the
maximal gain.

We further note that work of~\cite{Raj02,RSA04} also identified that additional queuing delay can improve the outage
performance. The change of slope of the outage as a function of SNR was observed, though no results regarding actual increase in
diversity level were proven. In this paper, we show that queuing delay does increase diversity for all multiplexing gains by
simultaneously using rate and power control.

The fact that diversity order of the system can be increased beyond the available degrees of freedom is rather
counter-intuitive. Part of the elegance of the results in~\cite{ZT03} is that the diversity-multiplexing tradeoff curve
completely matches the intuition behind degrees of freedom framework, where each additional d.o.f.\ can be used to increase
either throughput or diversity. The degrees of freedom framework is a statistical notion, where the degrees are related to the
probability distribution of the channel. When the transmitter has no channel state information, it has to use the \emph{same}
action for each state and has no opportunity for adaptation.

However, with channel state information, the transmitter has the opportunity to adapt to current channel conditions. Thus, the
transmitter can adapt \emph{how} it uses the instantaneous d.o.f. This per state adaptation is key to changing the probability
of outage events, saving power in the good states to put more power in the poor channel states and thus increasing the number of
states in which a successful transmission occurs.

The rest of the paper is organized as follows. In Section~\ref{se:problem}, we introduce our channel and feedback model. In
Section~\ref{se:perfect}, we provide the results pertaining to the case of perfect channel information at the transmitter, while
Section~\ref{ref:quantize} and Section~\ref{sec:div-mux} have the quantized power control and diversity-multiplexing tradeoff
bounds, respectively, for finite bit feedback model. Finally, a joint rate and power extension is studied in
Section~\ref{sec:discuss}, where we also discuss the system design implications of our proposed power control method. We
conclude in Section~\ref{se:conclude}.

\section{Problem Formulation\label{se:problem}}

In this section, we first describe the channel and feedback system model, followed by the optimization objective of feedback
design.

\subsection{Channel Model}

\begin{figure}[hbtp]
\centering \epsfysize=2in \epsffile{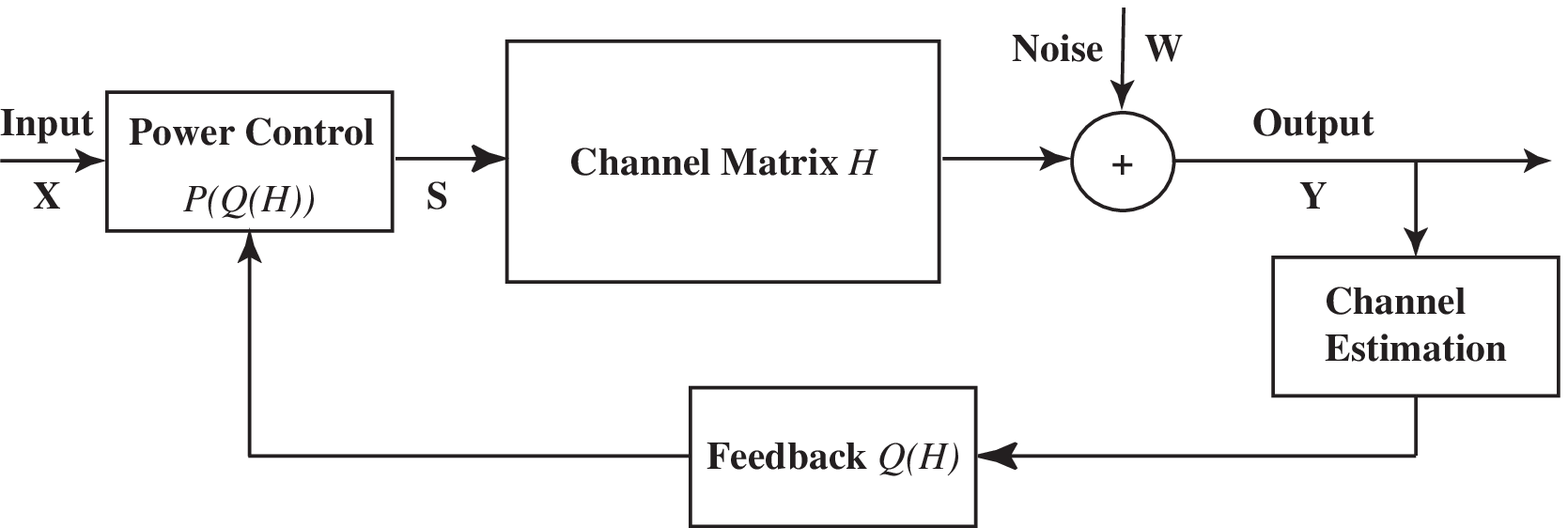} \caption{System with feedback.} \label{fig-fm}
\end{figure}

We consider a multiple antenna channel with $M$ transmit and $N$ receive antennas (Figure~\ref{fig-fm}). When the channel
coherence time is larger than the duration of transmission of a codeword, channel can be modeled as a block fading channel with
the following input-output relation
\begin{equation}
Y = H S + W\label{eq-mod1}.
\end{equation}
In~\ref{eq-mod1}, $S$ is $M\times 1$ transmitted vector, $H$ is the $N\times M$ channel matrix, and $W$ is the $N\times 1$
additive white Gaussian noise (AWGN). Channel matrix is assumed to have entries $h_{ij}$, that are scalar channel coefficients
between the $j^{th}$ transmit and $i^{th}$ receive antenna. Channel coefficients are assumed to be independent complex random
variables with circularly symmetric Gaussian distribution with zero mean and unit variance, $[h_{ij}]\in \mathcal{CN}^{N\times
M}$. The input signal vector $S\sim\mathcal{CN}^M(0,\Sigma_S)$ and the additive noise
$W\sim\mathcal{CN}^N(0,I_N)$.\footnote{$I_t$ is the $t\times t$ identity matrix.} Finally, the transmitter is equipped with a
finite average power $\p$ such that $\mathbb{E}[S^\dagger S]\le \p$.

\subsection{System Model}

The complete feedback system is depicted in Figure~\ref{fig-fm}, where there is fast feedback link between the transmitter and
receiver. The receiver is assumed to have perfect knowledge of the MIMO channel $H$, and sends a feedback codeword $Q(H)$ to the
transmitter. For the case of perfect information at transmitter $Q(H) = H$ and for finite rate feedback $Q(H): {\mathbb C}^{MN}
\rightarrow \{1,\ldots,L\}$. On receiving the feedback codeword $Q(H)$, the transmitter uses the channel information to adapt
its transmission scheme for actual transmission of the data. We assume that the feedback is error-free and delay-free. In our
recent work, we have also analyzed the impact of feedback errors for a two-way training model~\cite{SKSA05}.

The finite bits of feedback about the channel can be used to perform either beamforming~\cite{NLTW98,MSEA03,LHS03}, power
control~\cite{KS04,KS05} or rate control~\cite{LYS03,KS04a} among other possible transmission adaptations. In this paper, we
will focus mainly on temporal  power control (with a discussion of rate control in Section~\ref{sec:discuss}). In temporal power
control, all transmit antennas are used simultaneously with the same power $P(Q(H))$ or more concisely $P_Q(H)$. In this case,
the transmitted signal $S = P_Q(H)^{1/2} X$ and the received signal can be written as
\begin{equation}
  Y = H P_Q(H)^{1/2} X + W. \label{eq-mod}
\end{equation}
Note that $X$ has a unit power, i.e., $\mbox{tr}\left\{ {\mathbb{E}[XX^\dagger]} \right\} = 1$. If $X$ has iid components, then
$\mathbb{E}[XX^\dagger]=I_M/M$. The only constraint imposed is on the long term average power constraint of the transmitted
signal $P(Q(H))^{1/2} X$. That is,
\begin{equation}
\lim_{T\rightarrow\infty} \frac{1}{T}\sum_{i=1}^T \left[ P_Q(H_i)^{1/2}X_i)\right]^\dagger P_Q(H_i)^{1/2}X_i\le
P_{av}\label{eq-tpc1},
\end{equation}
where $H_i$ and $X_i$ represent the channel and the input signal at time $i$, respectively. Assuming a stationary and ergodic
channel and applying the law of large numbers, we can replace the time average with the ensemble average, i.e.,
\begin{eqnarray}
\lim_{T\rightarrow\infty} \frac{1}{T}\sum_{i=1}^T \left[ P_Q(H_i)^{1/2}X_i)\right]^\dagger P_Q(H_i)^{1/2}X_i &=& \mathbb{E}[X^\dagger P_Q(H) X]\nonumber\\
&\stackrel{a}{=}& \mathbb{E} \left[\text{tr}\left\{P_Q(H) XX^\dagger\right\}\right]\nonumber\\
&\stackrel{b}{=}& \frac{1}{M}\text{tr}\left\{\mathbb{E}\left[P_Q(H)\right]\right\}\label{eq-erg},
\end{eqnarray}
where (a) is obtained by applying $ x^\dagger y = \text{tr} \{y x^\dagger\},\;\forall x,y$ in a vector space $V$, and (b) by
having input signal with unit energy, i.e., $\mathbb{E}[XX^\dagger]=\frac{1}{M}I_M$. The expectation in~(\ref{eq-erg}) is taken
with respect to the distribution of $H$.

\subsection{Objective}

Our optimization objective is to attain maximum diversity order $d$ for a fixed multiplexing gain of $r$. The optimum power
allocation is known for perfect CSIT, therefore, the optimal diversity-multiplexing curve can be determined. However, the
optimal power control with finite feedback is not known, thus our result is a lower bound for the optimal diversity-multiplexing
tradeoff. The diversity order of any transmission scheme is defined as the slope at which the probability of outage decays as
the function of SNR or equivalently transmit power $\pa$. More formally, it is defined as~\cite{ZT03}
\begin{equation}
d = \lim_{\p\rightarrow\infty}-\frac{\log(\ou)}{\log(\p)}.\label{eq-defd}.
\end{equation}
where $\ou$ is the probability of outage with power $\p$ and rate $R(\p)$. Note that outage is typically understood as an event
in which the transmitter sends a codeword and receiver is unable to decode it correctly \cite{OSW94}. However, with channel
state information, the transmitter may choose not to send any codeword in poor channel conditions. In the conventional
definition, such an action implies no outage, leading to potentially no outage system. We believe for feedback based system, a
more general definition is necessary. We visualize the system as a time slotted system, in which each slot can accommodate a
full codeword and the channel state is constant over the whole time-slot. The outage event is then defined as \emph{lack of
information} at the receiver at end of each time-slot. Thus the outage event can occur in two possible ways - either the
transmitter sends no information or the transmission is in error. In either case, the receiver gets no useful information. We
note that this definition is useful when the system is not allowed to queue the packets and each packet is either sent or
dropped. If queuing is allowed at the transmitter, then the above outage definition still applies as either a packet drop by the
transmitter or erroneous decoding at the receiver~\cite{RSA04}.

The complementary metric of multiplexing gain is defined as the rate at which throughput scales as a function of $\log(\pa)$,
computed as
\begin{equation}
r = \lim_{\p\rightarrow\infty}\frac{R(\p)}{\log(\p)}.\label{eq-defd}
\end{equation}
We note that multiplexing gain, $r$, can not be larger than the rate of growth of ergodic capacity, $m=\min(M,N)$ with any
amount of channel information at either transmitter or receiver.

\section{Perfect Transmitter and Receiver Information}\label{se:perfect}

In this section, we will derive the diversity-multiplexing tradeoff curve for multiple antenna channels with perfect channel
information at both the transmitter and receiver (CSITR). A complementary analysis with only receiver information was first
derived in~\cite{ZT03}. It is apparent that receiver only analysis is a lower bound to our finite bit feedback analysis in the
next section, and the perfect CSITR analysis in this section yields an upper bound. The transmitter channel information is
predominantly employed to perform power control, rate control and/or transmitter beamforming. In this section, we will restrict
our attention to methods which no \emph{not} perform rate control, and always send at a constant rate.

Both power control and beamforming are examples of adaptive power allocation. In the case of beamforming, only spatial power
adaptation is performed while each codeblock receives the same total power and thus there is no temporal adaptation of the total
power given to the codewords. In this section, we will compare the asymptotic performance of three methods. The first is the
optimal power control which performs a joint temporal and spatial power control, and was derived in~\cite{CTB99,BCT01}. Thus,
optimal power control implicitly performs beamforming. The second is purely temporal power control with equal power allocation
on each spatial direction.  Temporal power control does not require information about eigen-directions and thus needs lesser
feedback about the channel state information than optimal power control. Finally, we will consider spatial power control,
equivalently beamforming, and thus compare the efficacy of temporal versus spatial opportunism.

First, we state the diversity-multiplexing tradeoff with \emph{no} transmitter information and perfect receiver information as
derived in~\cite{ZT03}.
\begin{thm}[Theorem~2, \cite{ZT03}]
Assume that the codeword length $l > m + n -1$. The optimal diversity-multiplexing tradeoff curve $d^*(r)$ is given by the
piecewise linear function connecting the points $(r,d^*(r))$, $k=0,1,m$ where
\begin{equation}
d^*(r) = (m-r)(n-r)
\end{equation}
Recall  $n = \max(M,N)$ and $m = \min(M,N)$. \label{th:noCSIT}
\end{thm}

The diversity multiplexing curve captures the intuitive tradeoff in using the spatial degrees of freedom to increase data rates
or reliability or a combination of the two objectives. Without any instantaneous channel information at the transmitter, the
transmitter cannot perform any rate, spatial or temporal power control and uses all its power in all spatial directions for all
the channel conditions. In the next three theorems, we will explore the efficacy of power control and compare it with
Theorem~\ref{th:noCSIT}.
\begin{thm}[Optimal Power Control]
With the perfect channel information at both transmitter and receiver, the diversity order $d^\dagger(r)$ of optimal power
control for multiplexing gain $r$ is given as
\begin{eqnarray}
d^\dagger(r) = \begin{cases}
\infty & r \in [0,m) \\
\infty & r = m \mbox{ and } n > 2m.
\end{cases}.
\end{eqnarray}\label{the-csirt}
\end{thm}
\noindent {\bf Proof}: See Appendix~\ref{test}.

The Theorem~\ref{the-csirt} does not include the case when $m\le n\le 2m$ and $r=m$. We conjecture that when $r=m=n$ the
diversity order is zero. The intuition is based on the optimum power control in \cite{CTB99} is which the optimal power control
depends on the geometric mean of eigenvalues. Whether the channel parameter is the geometric mean of eigenvalues of any additive
or multiplicative function of eigenvalues, in either case the distribution of the parameter is of form $f_\eta(x) = \kappa
x^{n-m} e^{-x} q(x)$, \footnote{It is not relevant to the discussion in this paper, but the distribution of
$\bar{\lambda}=\left(\prod \lambda_i\right)^{1/m}$ can be found by classical method of finding the distribution of $z =
\log\bar{\lambda}$ and it can be shown that the roots of the distribution at origin has the multiplicity of (n-m). Similarly
through the convolution argument it can be shown that the distribution of the summation of arbitrary number of eigenvalues
including the smallest has roots at origin with multiplicity of no larger than (n-m).} where $\kappa$ is a normalizing constant
and $q(x)$ is a polynomial such that $q(0)\ne 0$. Therefore, for $n=m$, the average power constraint $\int_{\gamma_0}^\infty 1/x
f_\eta(x) dx = 1$ has a singularity at $x = 0$. That means that $\gamma_0\nrightarrow 0$. In fact $\gamma_0 = c$ for some
constant $c$, which in turn implies that the outage probability, $\Pr\{\eta < \gamma_0\}$ is constant. Similarly, when $n>m$,
the channel inversion resolved by roots of channel distribution at origin results in outage free power allocation for all
channel conditions. Thus, the outage probability is zero and diversity order is infinite.

When $r\ne m$ but $n=m$, $\gamma_0$ is still a constant but a function of $\p$ such that as $\p$ approaches infinity, $\gamma_0$
goes to zero. Thus, a non-zero diversity order is achievable. On the other hand, when $n>m$ the distribution of the channel
parameter has a root at the origin, that is, $f_\eta(0) = 0$, which cancels out the singularity of the channel inversion.
Therefore, it is possible to make $\gamma_0$ small and yet not violate the average power constraint. The challenge at $r=m$ is
that the equation for average power constraint is independent of $\p$ and hence $\gamma_0$ is a constant. If this constant is
zero, then the diversity order is infinity, but if the constant is non-zero, then the outage is constant and the diversity order
is zero. The complete solution of $\gamma_0$ depends on the constant $\kappa$ and the polynomial $q(x)$, which make the analysis
involved.

In general the behavior of systems at highest multiplexing gain $r = m$ is hardest to predict. The rate is growing with SNR as
fast as the ergodic capacity does, while the block length is kept fixed. Next theorem shows that when CSI is only available at
the receiver, the diversity order at r = m is not achievable.
\begin{thm}
At highest multiplexing gain $r=m$, the outage probability $\Pi_{CSIR}\nrightarrow 0$ as $\p\rightarrow\infty$. Thus, $r-m$ is
not achievable.\label{the-maxmux}
\end{thm}
{\bf Proof}: See appendix~\ref{app-p1}.\\

A comparison of Theorems~\ref{th:noCSIT} and \ref{the-csirt} shows that the transmitter information completely changes the
asymptotic decay rate of probability of outage from a finite decay in perfect CSIR system to exponential rate of decay in
perfect CSIRT system. The change in decay rate can be completely attributed to optimum power control, which was derived
in~\cite{BCT01} and stated below with transmission rate as $r \log(\s)$
\begin{eqnarray}
P_i = \left[\left(\frac{2^\frac{r \log\s}{m}}{(\prod_{j=1}^m\lambda_j)^{\frac{1}{m}}} \right) -
\frac{1}{\lambda_i}\right]^+.\label{eq-2}
\end{eqnarray}
The power $P_i$ is the power allocated to the eigenvector corresponding to eigenvalue $\lambda_i$. Thus, not only there is
spatial power control, the total power allocated for a codeword depends on the current channel $H$ leading to temporal power
control. In the next theorem, we allocate equal power to each antenna (no spatial power control) which can change from one
time-slot to another based on current channel conditions, i.e., temporal power control.

\begin{thm}[Temporal Power Control]
When temporal power control is performed with same power on all antennas, then the diversity order $d_t(r)$ is given as
\begin{equation}
d_t(r) = \begin{cases} \infty & r< m\\
0 & r = m\;,\; n<2m\\
\infty & r = m\;,\;n\ge 2m \\
\end{cases}
\end{equation}
\end{thm}
{\bf Proof:} Consider a power allocation policy based on the smallest eigenvalue $\lambda_m$. It is well known that the
distribution of the smallest eigenvalue is given by \cite{Ede89}
\begin{equation}
f_{\lambda_m}(x) = \frac{1}{\Gamma(n-m+1)}x^{n-m}e^{-x},\label{eq-3}
\end{equation}

The outage minimizing power allocation \cite{CTB99} is of the form
\begin{equation}
P(\lambda_m) = \begin{cases} \frac{2^R-1}{\lambda_m},\;\;\; &\lambda_m>\gamma_0\\
0, &\text{otherwise}
\end{cases}
\end{equation}
where $\gamma_0$ is the boundary of the outage region and is obtained from the relation for average power constraint, i.e.,
\begin{equation}
\int_{\gamma_0}^\infty m P(x) f_{\lambda_m}(x) dx \le \p.\label{eq-4}
\end{equation}
Then, the information outage occurs only when there is no transmission,
\begin{equation}
\Pi_{EP} = \Pr\{\lambda_m<\gamma_0\}.\label{eq-6}
\end{equation}
Replacing the distribution in (\ref{eq-4}) and solving for $\gamma_0$ we have
\begin{equation}
\int_{\gamma_0}^\infty m \frac{2^{\frac{r}{m}\log\p}-1}{\Gamma(n-m+1) x}x^{n-m} e^{-x} dx \le \p,
\end{equation}
which can be simplified to
\begin{equation}
\int_{\gamma_0}^\infty x^{n-m-1} e^{-x} dx \le \frac{\Gamma(n-m+1)}{m} \p^{1-r/m}.\label{eq-5}
\end{equation}
For $r<m$, the right hand side of (\ref{eq-5}) goes to infinity as $\p$ goes to infinity, whereas the left hand side of
(\ref{eq-5}) is bounded above by $\Gamma(n-m)$ when $\gamma_0 = 0$. Therefore, for $\p>\Gamma(n-m)*m/\Gamma(n-m+1)$, the cutoff
threshold $\gamma_0 = 0$ and the outage probability $\Pr\{\lambda_m<\gamma_0\}=0$, which results in an infinite diversity order.

For $r=m$, the right hand side of Equation (\ref{eq-5}) is independent of $\p$. Therefore, in order to have diversity order of
infinity, we need to have zero outage probability, which is equivalent of $\gamma_0 = 0$. For $\gamma_0 = 0$ inequality
(\ref{eq-5}) becomes
\begin{eqnarray*}
1 & \le & \frac{\Gamma(n-m+1)}{m\cdot\Gamma(n-m)}\\
& = & \frac{n-m}{m},
\end{eqnarray*}
or equivalently $n\ge 2m$. Therefore, for $n\ge 2m$ the outage probability is zero and diversity order of infinity is achieved
at $r=m$. However, for $n<2m$, $\gamma_0 = 0$ can not be the solution to (\ref{eq-4}) and therefore the outage would be
non-zero.\footnote{Note that the optimum choice of feedback is an open problem, and a better choice of feedback may change the
statement of the theorem.} \QED

Thus, the asymptotic behavior for all but maximum multiplexing gain is unchanged. In some cases (such as $n=m$), for the largest
multiplexing gain, the diversity order is reduced from $\infty$ to zero, much like in a system with no transmitter information.
The reason for such dramatic change is insistence on equal power allocation in all eigenvalue directions. In this case, the
diversity order at maximum multiplexing gain is determined by the minimum eigenvalue which leads to a zero diversity gain.

\begin{thm}[Spatial Power Control a.k.a.\ Beamforming]
When only spatial power control is performed with same total power for all  channel realizations, then the diversity order
$d_s(r)$ is same as the system with no transmitter information~\cite{ZT03}, and is given by a piecewise linear function between
the following points $(r,d_s(r))$ for $ r=0,1,\ldots,m$ where
\begin{equation}
d_s(r) = (m-r)(n-r) = d^*(r).
\end{equation}
\end{thm}
{\bf Proof:} The advantage of beamforming at the transmitter is in that signals at the receiver can be combined coherently and
hence the total received SNR is improved. The expression for mutual information is given by
\begin{equation*}
I(X;Y|H) = \log\left(\det I_N + \p HH^\dagger\right).
\end{equation*}
It is shown in \cite{MMSA02}, that all systems in which the expression for mutual information is of form $\log\det(I+\alpha \p
HH^\dagger)$, have the same diversity order. For a system with beamforming at the transmitter $\alpha = 1$, and for a channel
model with CSI only at the receiver $\alpha = 1/M$. Therefore, beamforming does not change the diversity order. \QED

The above theorem demonstrates that beamforming alone, with no temporal power control, leads to no improvement in asymptotic
rates. Thus, while beamforming leads to SNR improvement in outage by coherent combining of the received signals, it has no
impact on rate of decay of outage probability with $\s$~\cite{MMSA02}. The above sequence of results demonstrate that temporal
power control is better use of channel information than spatial power control. In other words, if there were only a finite bits
of information available at the transmitter about the channel, there impact may be highest if all of them are allocated to
performing temporal power control. \\

\noindent \emph{Example 1}: Consider the case of single-antenna system with $n = m = 1$. With perfect information about the
channel, the transmitter action includes a truncated power control and a phase correction at the transmitter. Thus, the received
signal is
\begin{eqnarray}
y &=& \begin{cases}
\left( e^{r\log\s} -1\right) s + w, & |h|^2\geq \gamma_0 \\
w, & |h|^2 < \gamma_0
\end{cases}.
\end{eqnarray}
The threshold $\gamma_0$ is chosen to meet the average power constraint. Note that since $\mathds{E}[1/|h|^2]=\infty$, then in
order to have a finite power constraint $\gamma_0>0,$ for all SNR. The outage probability is approximately $\Pi \approx
\exp\left( - \s^{1-r} \right)$ for $r \in [0, 1)$, which decays exponentially fast like in Gaussian channels.
Thus the diversity order is infinite for all multiplexing gains less than the maximum. \\

\noindent \emph{Example 2}: In a $1\times2$ system the channel distribution is slightly different. The channel norm $\|h\|^2$
distribution (with maximum ration combining at the receiver) is $f_{\|h\|^2}(x) = x e^{-x}$. Since, $f_{\|h\|^2}(0) = 0$, then
$\mathds{E}[1/\|h\|^2]\le c$ for some constant $c$. Thus, for $\p>c$ it is possible to invert all channel conditions and yet
meet the average power constraint. Therefore, for all multiplexing gain $r\in[0,1]$, a zero outage and equivalently infinite
diversity order is achievable. Note that unlike single
antenna case (\emph{Example 1}), diversity order of infinity is achieved at $r=1$.\\

For the rest of the paper, we will largely focus our attention on finite feedback based temporal power allocation since it
achieves large gains without the need for learning the eigenvectors of the channels (needed by optimal and spatial power control
methods).

\section{Quantized Power Control \label{ref:quantize}}

In this section, we will develop a suboptimal finite bit quantizer to perform temporal power control. The quantizer has a simple
form which allows recursive calculation of all quantization thresholds, and simplifies the subsequent analysis of
diversity-multiplexing tradeoff.

\subsection{Preliminaries}

In temporal power control, each eigen-direction receives the same power which varies from codeword to codeword based on current
channel conditions. In this case, the mutual information using a full-rank Gaussian space-time code with covariance $\frac{1}{M}
I_{M}$ and power $P(H)$  is given by
\begin{equation}
I(S;Y|Q(H)) = \log \det \left( I_N + \frac{P(H)}{M} H I_M H^\dagger \right), \label{eq:mi1}
\end{equation}
where $I_M$ is $M\times M$ identity matrix and $H^\dagger$ is the Hermitian conjugate of $H$. The power $P(H)$ is the equal
power assigned to all the eigenvectors and depends on channel conditions $H$. The mutual information in Equation~(\ref{eq:mi1})
can be rewritten as
\begin{eqnarray}
I(S;Y|Q(H))  &=&  \log \det \left( I_N + \frac{P(H)}{M} H H^\dagger \right), \nonumber \\
&\stackrel{(a)}{=}& \log \det \left( I_N +  \frac{P(H)}{M}  (U \Lambda^{1/2} V^\dagger )(U \Lambda^{1/2} V^\dagger)^\dagger \right), \nonumber \\
&=& \log \det \left( I_m + \frac{P(H)}{M} \Lambda \right), \nonumber \\
&=& \sum_{i=1}^m \log \left( 1 + \frac{P(H)}{M} \lambda_i \right),\label{eq-mut1}
\end{eqnarray}
where (a) is obtained by replacing $H$ with its singular value decomposition, $H = U \Lambda^{1/2} V^\dagger$ and $m =
\min(M,N)$. Furthermore, $\left\{ \lambda_i \right\}_{i=1}^m$ are the non-zero eigenvalues of the matrix $HH^\dagger$. Thus for
temporal control, the power $P(H)$ is a mapping from the $m$-dimensional eigenvector space $\begin{bmatrix} \lambda_1 &
\lambda_2  & \cdots & \lambda_m \end{bmatrix}$ to non-negative real space $\mathbb{R}^+ \cup \{ 0\}$, and is a complex
non-linear vector quantization problem.

Instead of using a vector quantizer, we will further simplify the quantization problem by focusing on only one of the $m$
eigenvalues. The power $P(H)$  is then determined by a single eigenvalue $\lambda_i$ and the functional relationship will be
denoted by $P(\lambda_i)$ whenever needed. The simplification reduces the $m$-dimensional vector quantization problem to a
single dimensional vector quantization. Next we consider the design of optimal scalar quantizer for a fixed eigenvalue.

\subsection{Optimal Single Eigenvalue Quantizer \label{sec:optimal}}

\begin{figure}[hbtp]
\centering \epsfysize=5in \epsffile{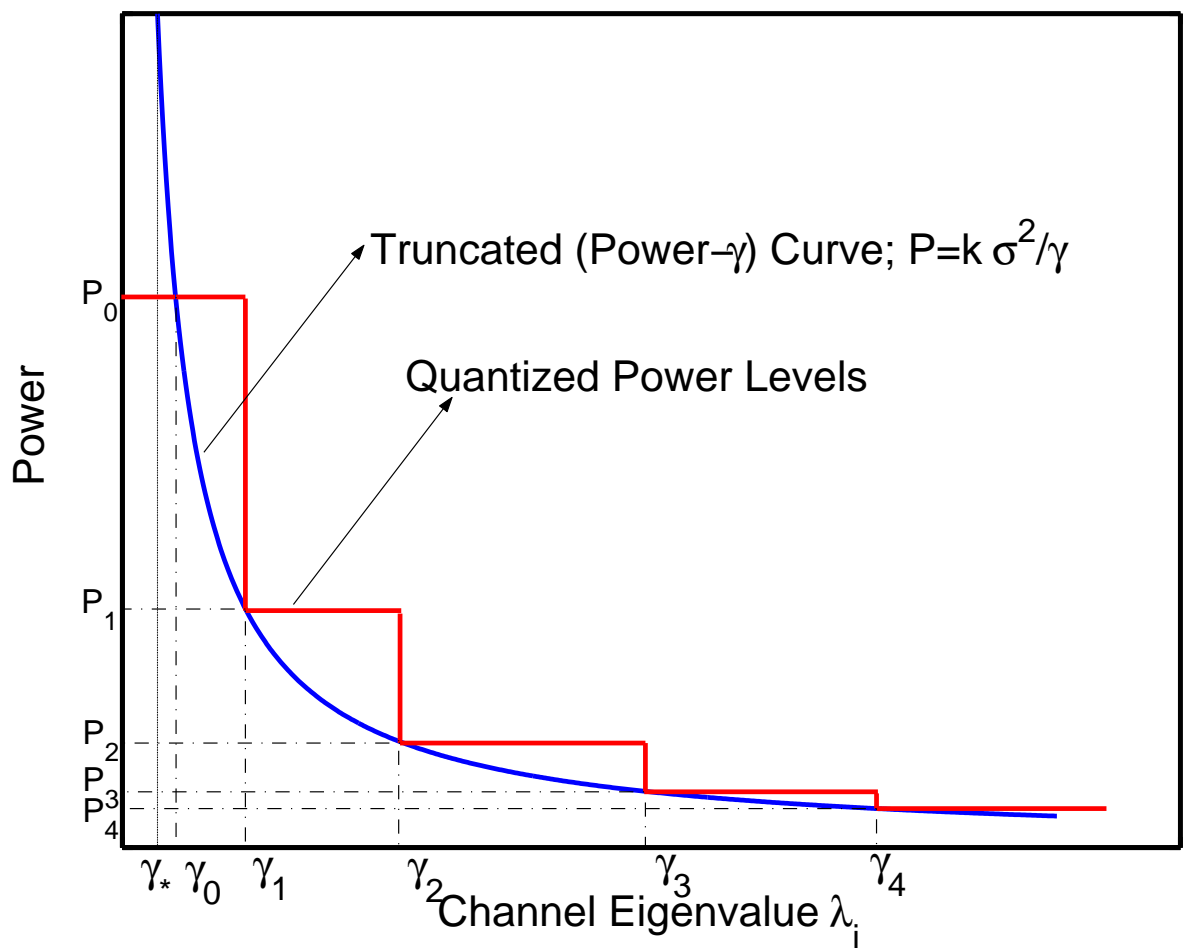} \caption{An example of a 5-bin channel quantizer with channel inversion
power allocation.} \label{fig-trunc}
\end{figure}

A typical continuous and quantized power control is illustrated in Figure~\ref{fig-trunc}. The continuous curve in
Figure~\ref{fig-trunc} corresponds to channel inversion power allocation, and the piecewise continuous step function is its
quantized approximation with $5$ quantization bins. The $x$-axis in Figure~\ref{fig-trunc} represents one of the non-zero
eigenvalues of the $HH^\dagger$ matrix, say $\lambda_i$.  If the  receiver observes that  the eigenvalue $\lambda_i$ lies in the
interval $[\gamma_2, \gamma_3)$, the index of this interval is fedback to the transmitter. Transmitter in turn allocates power
$P_2 = k/\gamma_2, k=(2^{R(P_{av})}-1)$. Since the transmitter does not know the exact value of the channel, it should allocate
power based on the worst case scenario such that for all channel conditions in that interval, an outage free communication is
guaranteed.

The channel quantizer is described by $L$ quantization thresholds $\{ \gamma_1, \gamma_2, \ldots, \gamma_{L-1} \}$ which define
$L$ quantization bins $\{ [0,\gamma_1), [\gamma_1, \gamma_2) , \ldots, [\gamma_{L-1}, \infty) \}$, where $L=2^B$. For each
quantization bin $[\gamma_j,\gamma_{j+1})$, $j \geq 1$, power $P_j = k/\gamma_j$ with $k = (2^{R(P_{av})}-1)$. Since origin
belongs to the first quantization bin, a finite power $P_0$ is assigned to this bin such that the total average power constraint
is satisfied. Note that $0 < (\gamma_0 = k/P_0 < \gamma_1$ such that power $P_0$ only guarantees outage free communication for
$\lambda_i \geq \gamma_0$. For $\lambda_i < \gamma_0$, the power $P_0$ is not sufficient to prevent outage and thus the
probability of outage is given by
\begin{equation}
\Pi(R(\p)) = \text{Prob} \left\{ \lambda_i < \gamma_0 \right\}.\label{eq-o1}
\end{equation}

First we note that the average power constraint of (\ref{eq-erg}) with quantized power levels is reduced to
\begin{equation}
{\mathbb E}[P^*(Q^*(\lambda_i))] = P_0 F_{\lambda_i}(0,\gamma_1) + \cdots + P_{L-1} F_{\lambda_i}(\gamma_{L-1},\infty)
\label{eq-ergq}
\end{equation}
where $F_{\lambda_i}(\alpha,\beta)~=\int_\alpha^\beta f_{\lambda_i}(x)dx$ is the probability mass concentrated in the interval
$[\alpha,\beta]$ and $f_{\lambda_i}(\cdot)$ is the probability distribution of $\lambda_i$.

Then the optimum channel quantizer ${Q^*}$ along with the optimum quantized power allocation $P^*(Q^*)$ are solutions to the
outage minimization problem
\begin{equation}
\{P^*(Q^*(\lambda_i)),Q^*(\lambda_i)\} = \arg \min_{\mathds{E}(P(Q(\lambda_i))) \leq \p} \Pi(R(P_{av})). \label{eq-opt1}
\end{equation}
In \cite{CTB99,BCT01} authors showed that the problem (\ref{eq-opt1}) has a dual which can be expressed by
\begin{equation}
\{P^*(Q^*(\lambda_i)),{\cal Q}^*(\lambda_i)\} = \arg \min_{\Pi(R(P_{av})) \leq \alpha} {\mathbb E}(P(Q(\lambda_i))).
\label{eq-dual}
\end{equation}

The constraint on outage in dual problem of (\ref{eq-dual}) is the same as saying $\Pr\{\lambda_i<\gamma_0\}\le\alpha$, which
can be solved for $\gamma_0$. Knowing $\gamma_0$, power level $P_0 = k/\gamma_0$ is known. Therefore the dual problem
(\ref{eq-dual}) is reduced to an unconstrained optimization problem in a space with one less dimension than the original
problem. The solution to the reduced optimization problem must satisfy the first order KKT condition, $\vec\nabla_{P(\lambda_i)}
{\mathbb E}_{\lambda_i}[P(\lambda_i)]=0$, which leads to the following system of nonlinear equations
\begin{equation}
\label{eq-soe}\begin{cases}
\frac{f_{\lambda_i}(\gamma_1)}{\gamma_0}-\frac{F_{\lambda_i}(\gamma_1,\gamma_2)}{\gamma_1^2}-\frac{f_{\lambda_i}(\gamma_1)}{\gamma_1} &= 0\\
\frac{f_{\lambda_i}(\gamma_2)}{\gamma_1}-\frac{F_{\lambda_i}(\gamma_2,\gamma_3)}{\gamma_2^2}-\frac{f_{\lambda_i}(\gamma_2)}{\gamma_2}&= 0 \\
\;\;\;\vdots \hspace*{2em} & \vdots\\
\frac{f_{\lambda_i}(\gamma_{L-1})}{\gamma_{L-2}}-\frac{F_{\lambda_i}(\gamma_{L-1},\infty)}{\gamma_{L-1}^2}-\frac{f_{\lambda_i}(\gamma_{L-1})}{\gamma_{L-1}}
&= 0.\end{cases}
\end{equation}
The solution to system of equations in (\ref{eq-soe}) does not admit a closed form and includes nonlinear transcendental
equations for Rayleigh channels. In next section we find a suboptimum channel quantizer which allocates equal total power to
each quantization bin, i.e., the product of power level and probability mass is equal across all quantization bins.

\subsection{Equi-Power Quantization \label{sec:equi-power}}

Consider the $j$th equation in (\ref{eq-soe}) ($1\le j\leq L-1$, with $\gamma_{L}=\infty$), that is,
\begin{equation}
\label{eq-ith}
\frac{1}{\gamma_{j-1}}f_{\lambda_i}(\gamma_{j})-\frac{1}{\gamma_{j}^2}F_{\lambda_i}(\gamma_{j},\gamma_{j+1})-\frac{1}{\gamma_j}f_{\lambda_i}(\gamma_j)=0.
\end{equation}
We can rewrite~(\ref{eq-ith}) as,
\begin{eqnarray}
\label{eq-rea}
\frac{1}{\gamma_j}(\gamma_{j+1}-\gamma_j)f_{\lambda_i}(\gamma_j)&=&\frac{1}{\gamma_{j+1}}F_{\lambda_i}(\gamma_{j+1}, \gamma_{j+2})\nonumber\\
P_j(\gamma_{j+1}-\gamma_j)f_{\lambda_i}(\gamma_j) &=& P_{j+1}F_{\lambda_i}(\gamma_{j+1}, \gamma_{j+2}).
\end{eqnarray}
As number of bits in feedback, $B=\log_2(L)$, approaches infinity, the length of quantization bins, $(\gamma_j, \gamma_{j+1})$,
approaches zero, and hence by mean value theorem~\cite{FF89}, we can further simplify~(\ref{eq-rea}) when $B\rightarrow\infty$
as
\begin{equation}
\label{eq-rea2} P_j F_{\lambda_i}(\gamma_j, \gamma_{j+1}) \approx P_{j+1}F_{\lambda_i}(\gamma_{j+1}, \gamma_{j+2}).
\end{equation}
The term $P_i F_{\lambda_i}(\gamma_j, \gamma_{j+1})$ is the total allocated power to the $j$th bin. Thus from~(\ref{eq-rea2}),
it follows that an approximation to the optimal power allocation is to allocate equal total power to each quantization bin. From
the above discussion, it also follows that the equal allocation power control is asymptotically (in number of quantization bins
$L$) approaches the optimum quantized power. The above approximate solution~(\ref{eq-rea2}) can now be used in the primal
problem~(\ref{eq-opt1}). Authors in \cite{CTB99} showed that the solution to~(\ref{eq-opt1}) is on the boundary of constraint
set, i.e., at the optimum point, $P^*(Q^*(\lambda_i))$, we have $\mathbb{E}[P^*(Q^*(\lambda_i))]=P_{av}$. More precisely, at
$P^*(Q^*(\lambda_i))$ we have
\begin{equation}
\label{eq-pav} P_0 F_{\lambda_i}(0,\gamma^*_1)+\dots+P_{L-1} F_{\lambda_i}(\gamma^*_{L-1},\infty) = P_{av}.
\end{equation}
Combining~(\ref{eq-pav}) with~(\ref{eq-rea2}) we get,
\begin{equation}
\label{eq-eqpower} P_j F_{\lambda_i}(\gamma^*_{j}, \gamma^*_{j+1})=\frac{P_{av}}{L}, \;\;\; \forall\;j\in\{0,2,\dots,L-1\},
\end{equation}
with $\gamma^*_L=\infty$ and $\gamma^*_0=0$.\footnote{to prevent more complication and only in here, we abuse the notation
$\gamma_0=0$ to indicate the left boundary of the left most quantization bin and every where else by $\gamma_0$ we mean the
threshold that determines the outage boundary and is equal to $k/P_0$} For $j=(L-1)$ in (\ref{eq-eqpower}) we have
\begin{equation}
\label{eq-lastbin} P_{L-1} F_{\lambda_i}(\gamma^*_{L-1}, \infty)=\frac{P_{av}}{L}.
\end{equation}
Also by channel inversion power allocation we have $\gamma^*_{L-1}~=~k/P_{L-1}$. Hence identity~(\ref{eq-lastbin}) is only a
function of $P_{L-1}$ (or $\gamma^*_{L-1}$) and we can solve~(\ref{eq-lastbin}) for $P_{L-1}$ (or $\gamma^*_{L-1}$). Replacing
the value for $\gamma^*_{L-1}$ in~(\ref{eq-eqpower}) for $j=(L-2)$, we end up with an equation with a single variable $P_{L-2}$
(and corresponding threshold $\gamma^*_{L-2}$). By recursively repeating the same procedure, we can obtain all the power levels,
$\{P_j\}_{j=1}^{L}$.

\begin{figure}[hbtp]
\centering \epsfysize=5in \epsffile{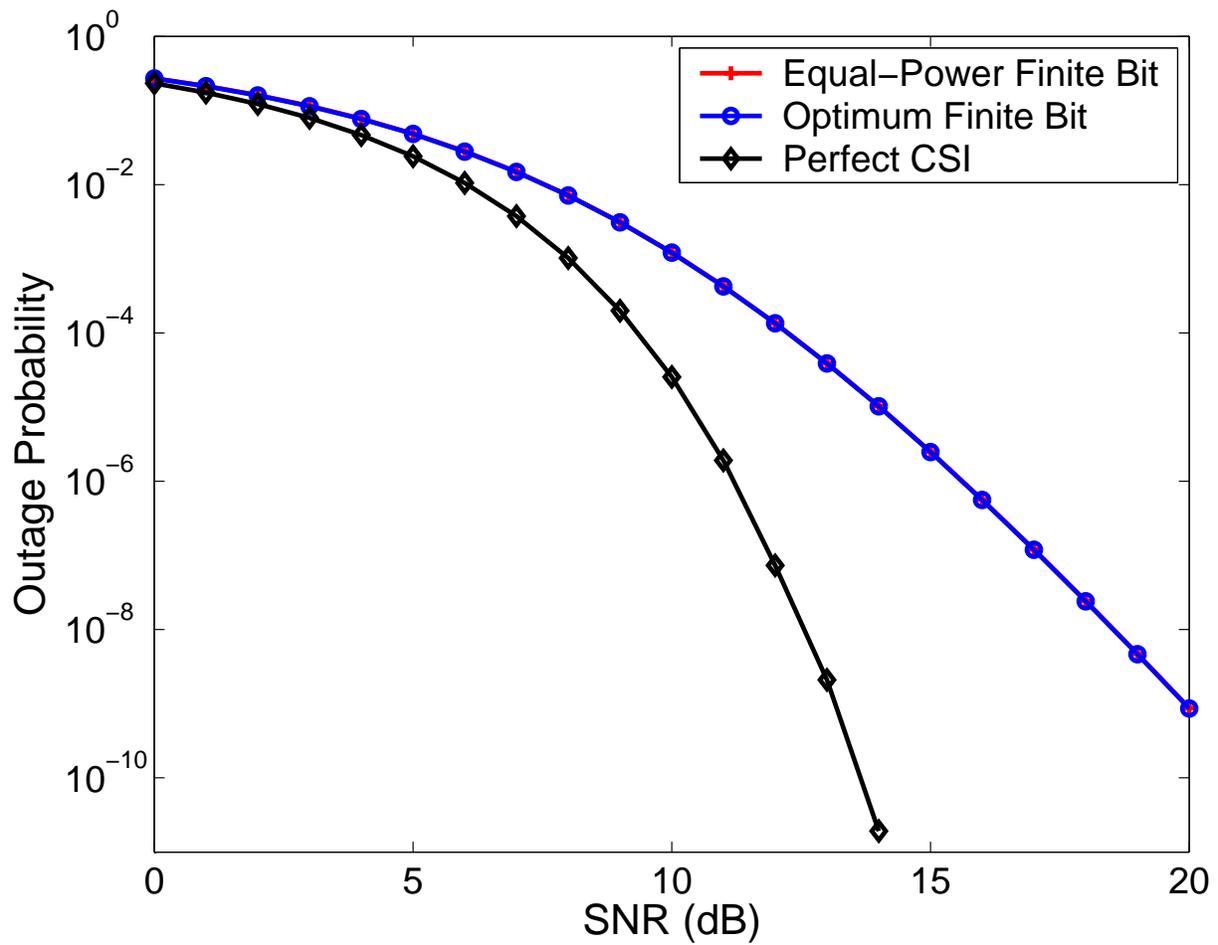} \caption{Outage for $\log_2(3)$ bits of feedback for optimal, equal
allocation power control, and perfect CSI at Tx and Rx with respect to SNR for a system with single antenna at transmitter and
receiver, and  transmission rate of R=2 b/s/Hz} \label{fig-ossnr}
\end{figure}

Figure~\ref{fig-ossnr} compares the performances of systems with quantized feedback, optimal, and equal allocation power
control, and a system with perfect CSI at both ends, as a function of SNR. A single transmit and single receive antenna system
is considered, and the feedback rate is $B=\log_2(3)$~bits/code-block. Note that the performance of optimal and equi-power
schemes are not distinguishable in Figure~\ref{fig-ossnr} indicating that equal power allocation performs very close to optimum
for the range of simulated SNRs.
%
%

Note that as $\p$ approaches infinity all the quantization thresholds $\gamma_i$'s, approach to zero. Therefore, the length of
all the quantization bins, $(\gamma_i, \gamma_{i+1}]$ approaches zero, which satisfies the mean value theorem condition applied
to (\ref{eq-rea}). Thus, the suboptimum quantizer found in this section is asymptotically optimum as $\p$ approaches infinity,
and the diversity order analysis based on the suboptimum power control would be the same with that of the optimum power control
based on the optimum quantizer.

The derivation of the quantization thresholds in this section is independent of the distribution of the channel parameter, as
long as the distribution is continuous and differentiable. In the next section we use the tools developed in here to
characterize the outage performance and quantify the diversity order of channel with Rayleigh distribution and finite rate
feedback.


\section{Diversity-Multiplexing Tradeoff with Quantized Feedback\label{sec:div-mux}}

The main idea in deriving the diversity-multiplexing tradeoff bound is as follows. To achieve a multiplexing gain of $i$, we
will use the largest $i$ eigenvalues and ensure that the allocated power is sufficient to achieve minimal outage. To do that we
will quantize $\lambda_i$ using the equi-power quantizer developed in Section~\ref{sec:equi-power}, which is designed to closely
approximate the optimal single-eigenvalue quantizer of Section~\ref{sec:optimal}. Having chosen the quantization mechanism based
on single eigenvalue, we then can study scalar eigenvalue distribution for high SNR regime which requires characterizing the
eigenvalue distribution around zero.
\begin{thm}[Diversity-Multiplexing Tradeoff]
In a MIMO system, for a multiplexing gain $r$, when a scalar quantizer with $L$ bins is used at the receiver the diversity order
is given by
\begin{equation}
d = \max_{i\in\{j,\dots,m\}}\left\{(1-\alpha)(n-j+1)(m-j+1)G(m,n,i,L)\right\}, \label{eq-dmax}
\end{equation}
where $j=\lceil r\rceil$, $\alpha = r/i$, and $i$ is the argument that maximizes (\ref{eq-dmax}). Further the function
$G(m,n,i,L)$ is defined as
\begin{equation}
G(m,n,i,L) \triangleq \sum_{l=0}^{L-1} [(n-i+1)(m-i+1)]^l.\label{eq-t1}
\end{equation}
\label{theo-thetheorem}
\end{thm}
{\bf Proof:} See Section~\ref{sec:proof}.
\QED\\

\noindent \emph{Example 3}: Consider the case of single transmit and receive antenna system, $m=n=1$. The diversity order for
zero multiplexing gain ($r=0, i=j=1$) is given by $d = 2^B$ (recall $L=2^B$). In contrast, with no CSIT, the diversity order is
one while with perfect CSIT, the diversity order is $\infty$. Thus, as expected, partial CSIT results in a finite diversity
order which increases with the number of bits. In other words, feedback adds to the diversity order of the system, primarily by
temporally utilizing spatial degrees of freedom. The more interesting aspect is the rate at which the diversity order grows; the
next example clarifies that further.\\

\noindent \emph{Example 4}: Consider the case of $m\times n$ system with $B$ bits of feedback. The diversity order for zero
multiplexing gain is $d = mn((mn)^{2^B}-1)/(mn-1)$. The above expression clearly shows that the diversity order is doubly
exponential in number of feedback bits $B$.\\

It is interesting to notice that for a SISO system with $B$ bits of feedback ($L=2^B$) and $r=0$, the diversity order is $d_L=L$
which is larger than diversity order with CSIR only, $d_{CSIR} = 1$. In fact it has a diversity order of a MISO/SIMO system with
$L$ antennas and CSI at receiver, which suggests that spatial diversity in CSIR systems can be replaced by \emph{feedback
diversity} in a feedback based system.

The diversity order in the statement of the Theorem~\ref{theo-thetheorem} consists of three parts; (i) the maximization, (ii)
the diversity order factor $(1-\alpha)(n-j+1)(m-j+1)$, (iii) and the effect of the quantization $G(m,n,i,L)$. The maximization
is simply finds the largest achievable diversity order, given the desired multiplexing gain. The factor $(1-\alpha, \alpha =
r/i$ is the result of dividing the desired rate equally among the at most $i=\lceil r\rceil$ many ``parallel" channels.
Therefore, $i$ indicates the smallest integer larger than the desired multiplexing gain. However, if the user decides to reduce
the multiplexing gain to a smaller value $0<j<i$, then the probability of the outage depends on the outage of the $j$ many
channels corresponding to $j$ largest eigenvalues, instead of $i$ many of them. The channel quantizer is fixed and transmitter
and receiver do not change the channel quantizer and power allocation. Thus, the effect of quantized power control $G(m,n,i,L)$
remains unchanged, even when the system operates at a lower multiplexing gain.

\begin{figure}[hbtp]
\centering \epsfysize=5in \epsffile{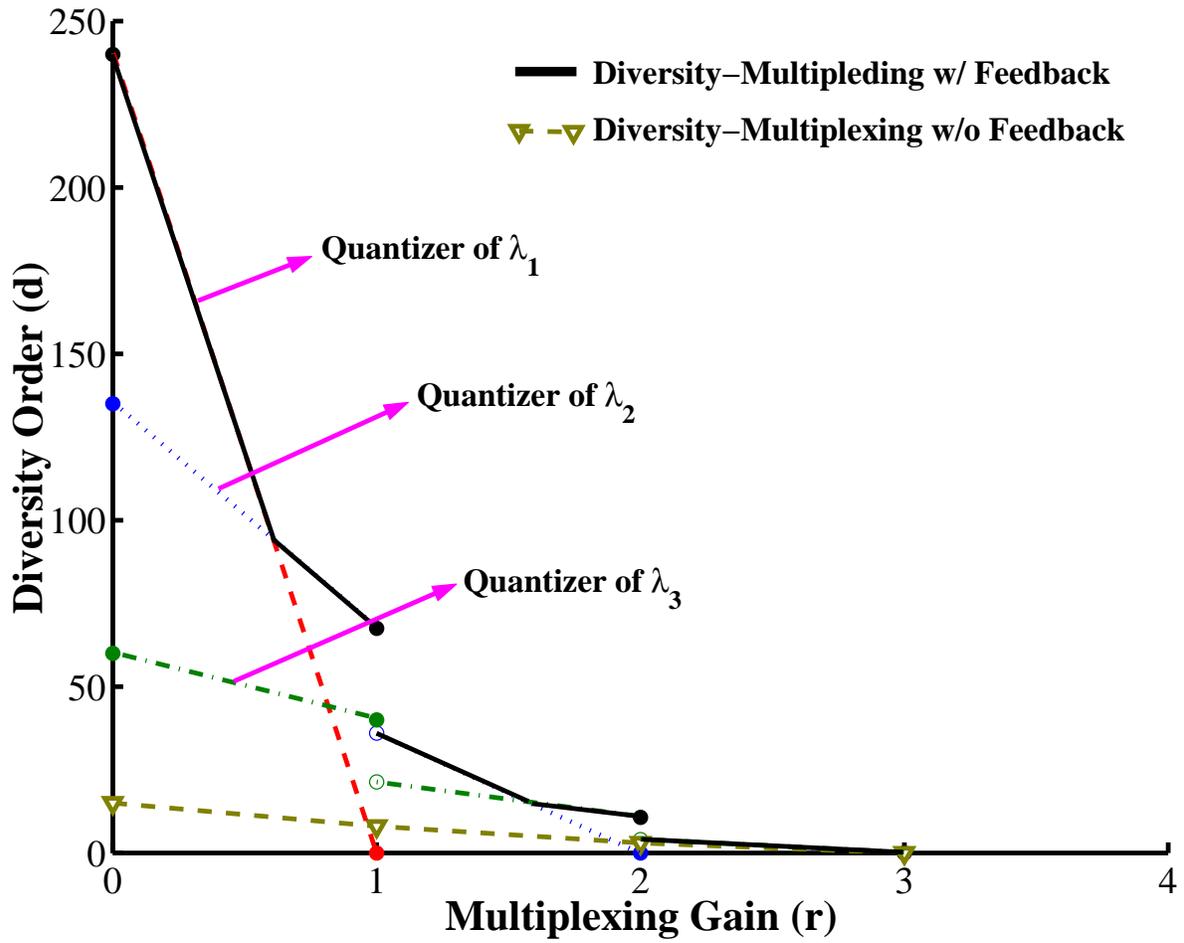} \caption{Diversity and Multiplexing curve with 1 bit of feedback and without
feedback as in \cite{ZT03}, for system with $m=3, n=5$.} \label{fig-rd}
\end{figure}

To elaborate the above discussion, Figure~\ref{fig-rd} depicts the diversity-multiplexing curve given in the
Theorem~\ref{theo-thetheorem} for a system with 1 bit of feedback, $m=3$, and $n=4$. Also for the sake of comparison the
diversity-multiplexing curve of a system without feedback, as given in \cite{ZT03}, is plotted. There are three sets of curves
in Figure~\ref{fig-rd}. The solid curve, dashed curves with circle markings, and dashed with triangle. The dashed curve with
triangle markings at the bottom is the diversity-multiplexing curve of a system without feedback. The solid envelope is the
maximum diversity-multiplexing, which is the $\max$ operation in (\ref{eq-t1}). The discontinuities at the integer multiplexing
gains are due to the switching the channel quantizer to a smaller eigenvalue due to the operation at a smaller multiplexing gain
for a fixed choice of channel quantizer. The set of dashed lines with circle markings can be divided into three curves. Since
$m=3$, there are three eigenvalues that can be used to design the channel quantizer. The curve with highest slope that crosses
the multiplexing axis at $r=1$ corresponds to the quantization of $\lambda_3$, which is the largest eigenvalue. The dotted curve
with smaller slope that crosses the multiplexing axis at $r=2$ and has a discontinuity at $r=1$ corresponds to the quantization
of $\lambda_2$, the second largest eigenvalue. When a multiplexing gain larger than 1 is desired, the outage is determined by
the realization of $\lambda_2$, whereas for multiplexing gains smaller than one it suffices to receive data reliably on the
channel corresponding to $\lambda_3$. Therefore, the outage behavior is determined by the realization of $\lambda_3$. An
analytical description of the above is given in (\ref{eq-newout}). Finally the dashed and dotted curve with the smallest slope
(still with circle marking) that crosses the multiplexing axis at $r=3$ corresponds to the quantization of $\lambda_1$, which is
the smallest eigenvalue. Similarly there are discontinuities at intermediate integer multiplexing gains of $r=1,2$.

Note that for any choice of eigenvalue $\lambda_i$ as channel parameter, for the multiplexing gain $r=i$, the diversity order is
zero. This is the side effect of our view to the channel and achievable rate. When $i^{th}$ largest eigenvalue is chosen to be
the channel parameter, the channel is viewed as virtually $i$ parallel channel. To achieve a multiplexing gain of $r=i$ is the
same as achieving a multiplexing gain of $1$ from each of the virtual channels. Achieving a multiplexing gain of $1$ on a
channel with degrees of freedom of $1$ is the same as achieving the ergodic capacity with finite block length, which is
impossible. Therefore, at $r=i$ probability of error does not vanish with SNR, and diversity order does not exceed zero.

As the number of bits in feedback increases, the envelope curve stretches vertically ({\em e.g.} example 4 is an example of the
effect of number of bits on the value of the diversity order). Hence, one can expect that as the number of feedback bits
approaches infinity, i.e., perfect channel state information at the transmitter, the diversity gain approaches infinity for any
arbitrary multiplexing gain, which complies with results in \cite{GCD05}.

%
%

\begin{figure}[hbtp]
\begin{center}
$\begin{array}{cc}
    \epsfxsize=3in \epsffile{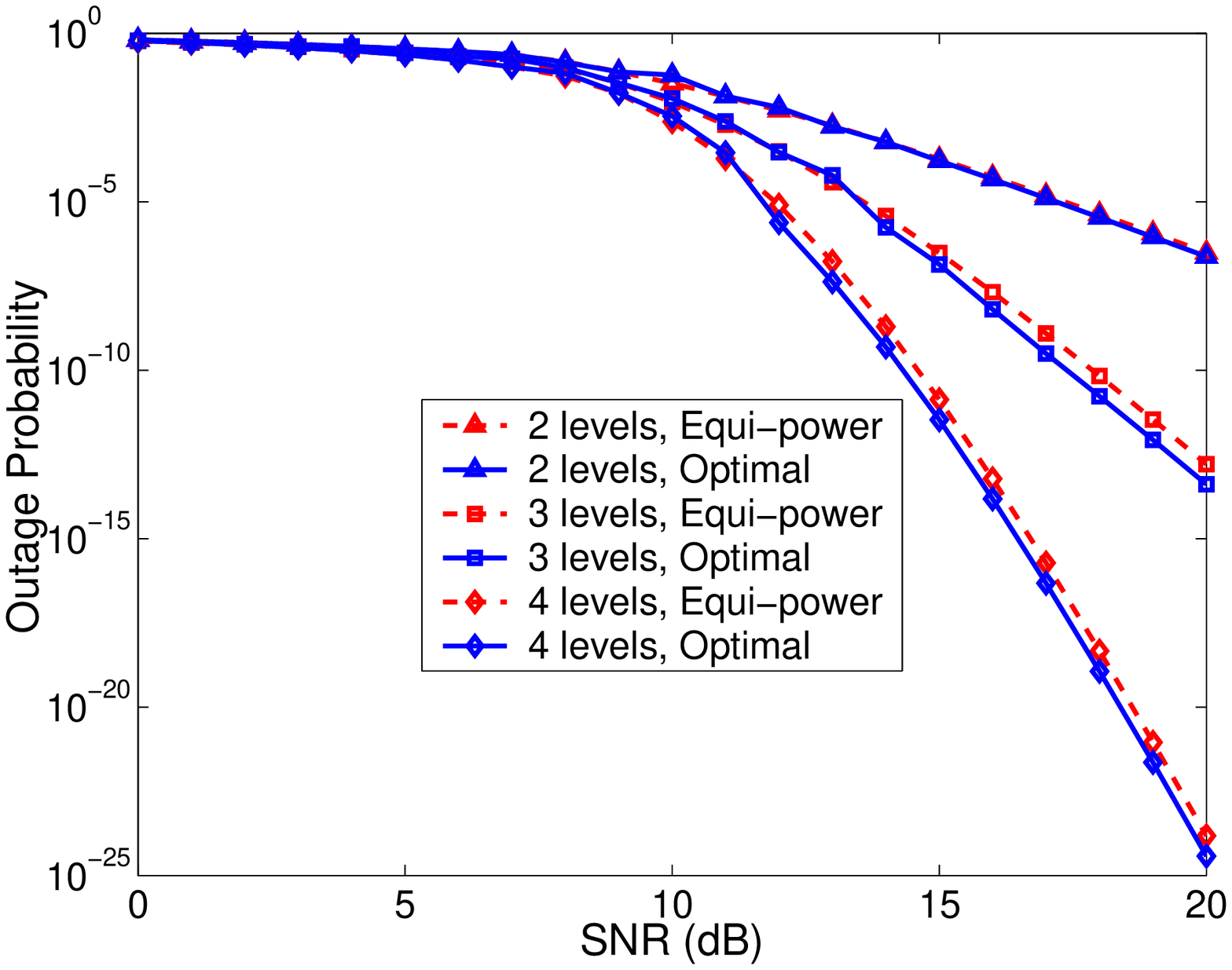} &
    \epsfxsize=3in \epsffile{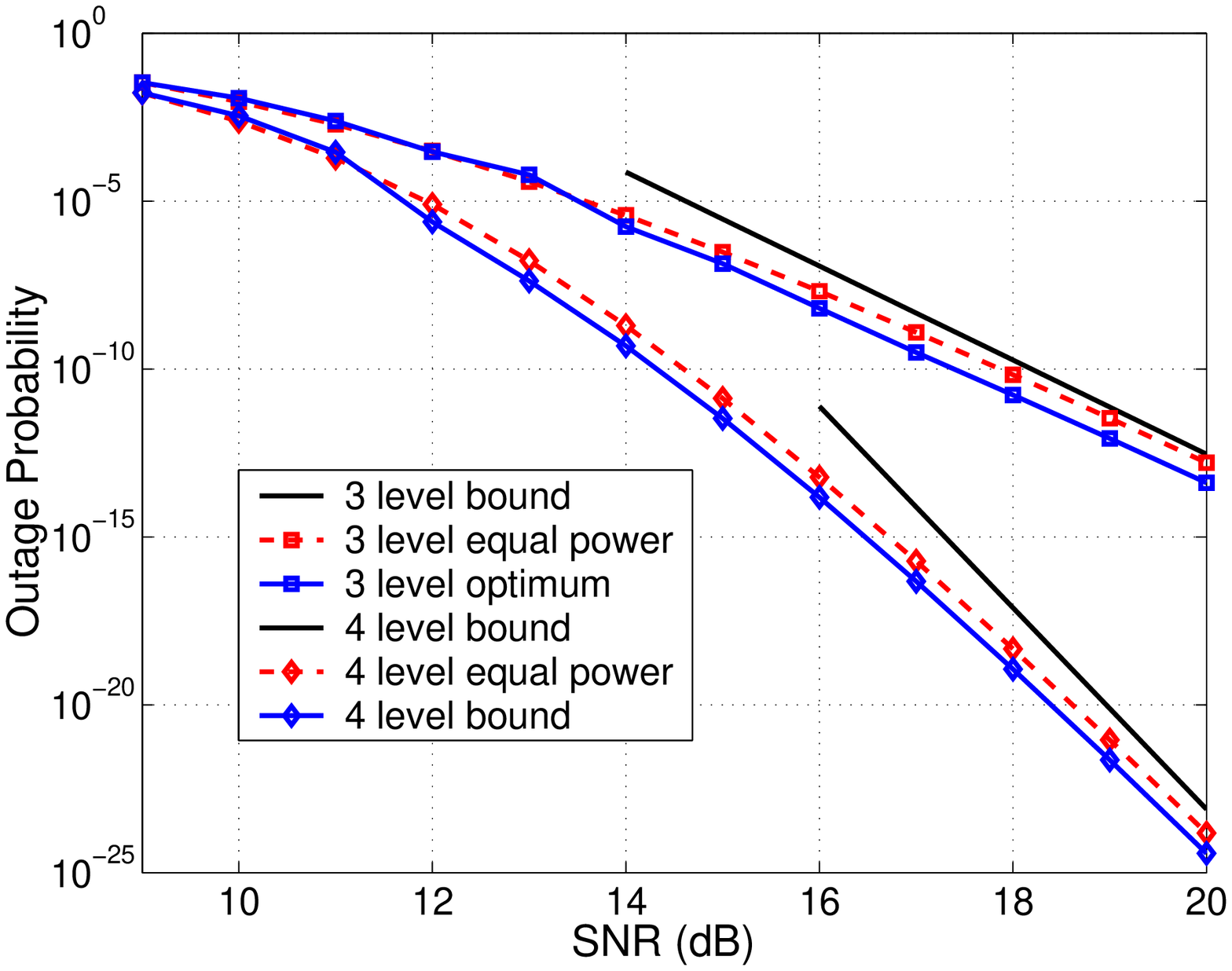}\\
    \mbox{(a)} & \mbox{(b)}
\end{array}$
\end{center}
\caption{(a) Outage probability as a function of SNR for a system with $L=2,3,4$, optimum solution of (\ref{eq-soe}) with
developed suboptimal recursive solution, (b) bound on diversity order in corollary~\ref{theo-thetheorem} for L=3, and 4 with
optimal and suboptimal power allocation schemes. Both graph are for a system with 2 transmit antennas, and rate R=2 bit/s/Hz.}
\label{fig-miso}
\end{figure}

Figure~\ref{fig-miso}~(a) shows the probability of outage of $2\times 1$ system with 1, $\log_2(3)$, and 2 bits of feedback
($L=2,3,4$) with constant data rate of 2 bits/s/Hz. The solid curves are the outage found from numerical solution of
(\ref{eq-soe}). The dashed curves are the outage calculated based on the developed suboptimal power control scheme. Even though
we showed the asymptotic optimality of our simplified scheme, Figure~\ref{fig-miso} also verifies our previous finding. Solid
lines in Figure~\ref{fig-miso}~(b) are diversity order for $L=3,4$ of the system with the same parameters as in (a).

\subsection{Proof of Theorem \ref{theo-thetheorem} ~\label{sec:proof}}

We divide the proof of Theorem \ref{theo-thetheorem} into three steps. Two Lemmas, which are interesting results by themselves,
provide the required tools for deriving the diversity order of quantized feedback. Body of the proof, which uses the results of
the lemmas and takes care of the subtleties to prove the statement of the theorem.

As we explained in the beginning of Section~\ref{sec:div-mux}, although conservative, in order to achieve multiplexing gain of
$r=i$, we allocate power that guarantees a multiplexing gain of 1 on the channel corresponding to $\lambda_i$. Similarly, we
define the outage, when the channel corresponding to the $\lambda_i$ is in outage, i.e., $\Pr\{\lambda_i<\gamma_0\}$. Therefore,
in order to find the outage probability, distribution of $\lambda_i$ is needed. In Lemma~\ref{lem-cdf} we find the distribution
of $\lambda_i, 1\le i\le m$.

The outage probability as given in (\ref{eq-o1}) depends on $\gamma_0$, which is determined by the quantization and power
allocation. Finding $\gamma_0$ requires solving the optimization problem (\ref{eq-opt1}). We find the quantizer and power
allocation using the developed equi-power quantizer. The function $G(\cdot,\cdot,\cdot,\cdot)$ in the statement of the theorem
is the effect of the quantized power control on the diversity order. In Lemma~\ref{lem-gg} we find the diversity order of
quantized feedback when $\lambda_i$ is being fedback to the transmitter.

The rest of the proof deals with the technicality of finding the maximum diversity order for a given multiplexing gain, which
explains having "max" in the statement of the Theorem~\ref{theo-thetheorem}. The argument is based on the idea of having the
larger eigenvalue as the channel parameter results in a smaller outage probability and hence larger diversity order.

The distribution of the ordered eigenvalues are unknown for the general case $\lambda_i$, and is only known for the smallest
eigenvalue~\cite{Ede89}. Since the diversity order is an asymptotic property, it suffices to know the asymptotic behavior of the
distribution of $\lambda_i$, which is provided in the next lemma.
\begin{lemma}[Asymptotic property of CDF of $\lambda_i$]
Let $\lambda_i$ be the $i$th largest eigenvalue with probability distribution function $f_{\lambda_i}(x)$. Define
$F_{\lambda_i}(t)$ by
\begin{equation}
F_{\lambda_i}(t) = \int_0^t f_{\lambda_i}(x)dx, \label{eq-fx}
\end{equation}
then the Taylor expansion of $F_{\lambda_i}(t)$ around the origin is given by
\begin{equation}
F_{\lambda_i}(t)=\beta_i t^{(n-i+1)(m-i+1)}+ o(t^l), \label{eq-fi}
\end{equation}
for some $l>(n-i+1)(m-i+1)$. \label{lem-cdf}
\end{lemma}
{\bf Proof:} Consider the ordered statistics of the eigenvalues of the matrix $Z$. The distribution of the ordered eigenvalues
of Wishart matrix, $\lambda_m > \lambda_{m-1} > \cdots
>\lambda_1>0$,  is given in \cite{Ede89}
\begin{equation}
f(\lambda_m,\lambda_{m-1},\dots,\lambda_1)=e^{-\sum_{i=1}^m \lambda_i} \prod_{i=1}^m \lambda_i^{n-m} \prod_{i<j}
(\lambda_i-\lambda_j)^2, \label{eq-joint}
\end{equation}
where $m=\min(M,N)$ and $n=\max(M,N)$.

While our analysis is based on the $i$th largest eigenvalue, for simplicity of the proof we study the asymptotic properties of
the $i$th smallest eigenvalue. The conversion can be done by change of indexing, that is the $i$th smallest eigenvalue is
$m-i+1$st largest.

Let the joint distribution of the $\lambda_i$'s, $1 \leq i \leq m$, be given by Equation~(\ref{eq-joint}). Then the marginal
distribution of $\lambda_i$'s, can be written as
\begin{eqnarray}
f_{\lambda_i}(t) &=\int_{\lambda_{m-1}}^\infty \int_{\lambda_{m-2}}^{\lambda_m} \cdots \int_{\lambda_i}^{\lambda_{i+2}}
\int_{\lambda_{i-2}}^{t} \cdots \int_{\lambda_1}^{\lambda_3}\int_0^{\lambda_{2}}
f(\lambda_m,\lambda_{m-1},\dots,\lambda_1)\nonumber\\
&d\lambda_1 d \lambda_{2} \dots d\lambda_{i-1} d\lambda_{i+1} \dots d\lambda_{m-1}d\lambda_m.\nonumber
\end{eqnarray}
Note that the integration over $\lambda_j$'s for $j>i$ results in a constant and is not a function of $\lambda_i$. On the other
hand, evaluation of $f_{\lambda_i}(t)$ involves $(i-1)$ integrations over $\lambda_j$'s, $1\le j<i$, each contributing a factor
of $\lambda_i^{n-m+1}$ to the marginal distribution.\footnote{$\int_0^x y^\alpha e^{-y}dy\approx y^{\alpha +1}$, for small $x$.}
Also for every $j$ and $k$ smaller than $i$, since $\lambda_i$ is assumed to be close to the origin, so are $\lambda_j$ and
$\lambda_k$ ($\lambda_{j,k}<\lambda_i$). Hence, we can approximate $(\lambda_k-\lambda_j)^2$ for $j<k<i$ with $\lambda_k^2$,
which results in a factor of $\lambda_i^{2(1+2+\dots+(i-1))}$. Therefore, $f_{\lambda_i}(t)$ is of form
\begin{equation}
f_{\lambda_i}(t)=t^{(n-m)+(i-1)(n-m+1)+2(1+2+\dots+(i-1))}q(t)e^{-t}, \label{eq-f}
\end{equation}
where $q(t)$ is a function of $t$ containing polynomials and exponential functions of $\lambda_i$ such that $q(0)\ne 0$. Let $k$
be the exponent of $x$ in Equation~(\ref{eq-f}). Since $\frac{dF_{\lambda_i}(t)}{dt}=f_{\lambda_i}(t)$, the first $k$
derivatives of $F_{\lambda_i}(t))$ evaluated at 0 are equal to 0. Thus, the Taylor expansion of $F_{\lambda_i}(t)$ around the
origin can be written as
\begin{equation}
F_{\lambda_i}(t)=\beta_i t^{k+1}+o(t^l),\label{eq-lem1}
\end{equation}
where $k+1 = i(n-m+i)$ and some $l>i(n-m+i)$. Note that from (\ref{eq-fx}) we have $F_{\lambda_i}(0) = 0$. Therefore, there is
no constant term in (\ref{eq-lem1}). By change of variable of $i$ to $m-i+1$ in Equation~(\ref{eq-lem1}) we have the expression
for the $i$th largest eigenvalue
\begin{equation*}
F_{\lambda_i}(t)=\beta_i t^{(n-i+1)(m-i+1)}+ o(t^l),
\end{equation*}
for some $l>(n-i+1)(m-i+1)$. \QED

As $\p$ increases, in order to meet the average power constraint (\ref{eq-ergq}), $P_0$ must increases, or equivalently
$\gamma_0$ should decrease. Therefore, as $\p$ approaches infinity, $\gamma_0$ approaches zero. Using the Lemma \ref{lem-cdf},
the outage expression (\ref{eq-o1}) as $\p\rightarrow\infty$ can be reduced to
\begin{equation}
\ou \approx \begin{cases} \gamma_0^{(n-i+1)(m-i+1)}\;\;\;\;\;\;&\text{MIMO}\\
\gamma_0^n &\text{MISO/SIMO}\\
\gamma_0 &\text{SISO}.
\end{cases}\label{eq-cor1}
\end{equation}

To quantify the diversity order of power control with finite rate feedback, we need to formulate the probability of outage as a
function of $\p$. A closer look at (\ref{eq-cor1}) one can notice that $\gamma_0$ is related to power allocated to the first
quantization bin $P_0$, and $P_0$ in turn is related to the $\p$ through the average power constraint (\ref{eq-ergq}). Using the
developed equi-power quantizer, we need to solve $L$ nonlinear transcendental equations of form (\ref{eq-eqpower}) in order to
find a closed form solution for $\gamma_0$. At large $\p$, a tight and accurate approximation to the equations
(\ref{eq-eqpower}) makes it possible to find $\gamma_0$ as a function of $\p$. The derivation is explained in next lemma.
\begin{lemma}
Consider a quantized power control with $L$ power levels which quantizes the $i$th largest eigenvalue of channel matrix. Define
\begin{equation}
G(m,n,i,L) \triangleq \sum_{l=0}^{L-1} [(n-i+1)(m-i+1)]^l.\nonumber \label{eq-G}
\end{equation}
Then for large values of $P_{av}$ the outage threshold $\gamma_{0}$ is given by
\begin{equation}
\gamma_{0} \approx \frac{c_i}{P_{av}^{(1-\frac{r}{i})G(m,n,i,L)}}, \label{eq-g0i}
\end{equation}\label{lem-g}
for multiplexing gain $r$, $0\le r\le i$, and some constant $c_i$.\label{lem-gg}
\end{lemma}
{\bf Proof:} Analyzing the suboptimum quantizer, with recursive equations explained in Section \ref{sec:equi-power}, with equal
total power at each quantization bin, and beginning from the equation for the last quantization bin (\ref{eq-eqpower})we have
\begin{eqnarray}
\frac{P_{av}}{L} &=& P_{L-1} \Pr\{\gamma_{L-1}<\lambda_i<\infty\}\nonumber\\
&\stackrel{a}{=}& \frac{\p^{r/i}}{\gamma_{L-1}}[1 -F_{\lambda_i}(\gamma_{L-1})]\label{eq-last}
\end{eqnarray}
where (a) is obtained by replacing power allocated to the $L^{th}$ bin, $P_{L-1}$, by its channel inversion equivalent. Solving
(\ref{eq-last}) we have
\begin{equation}
\gamma_{L-1} \approx \frac{c_{L-1}}{P_{av}^{1-r/i}} \label{eq-L}
\end{equation}
Note that having $\gamma_{L-1}$ from (\ref{eq-L}), the expression of power allocated at the $L-1^{st}$ bin given by
\begin{eqnarray}
\frac{P_{av}}{L} &=& P_{L-2} \Pr\{\gamma_{L-2}<\lambda_i<\gamma_{L-1}\}\nonumber\\
&=& \frac{\p^{r/i}}{\gamma_{L-2}}\left[F_{\lambda_i}(\gamma_{L-1})-F_{\lambda_i}(\gamma_{L-2})\right]\nonumber\\
&\stackrel{a}{\approx}& \frac{\p^{r/i}}{\gamma_{L-2}} F_{\lambda_i}(\gamma_{L-1})\nonumber\\
&\stackrel{b}{=}& \frac{\p^{r/i}}{\gamma_{L-2}} \left(\frac{c_{L-1}}{\p^{1-r/i}}\right)^{{(n-i+1)(m-i+1)}},\label{eq-l1}
\end{eqnarray}
where (a) is obtained by neglecting $F_{\lambda_i}(\gamma_{L-2})$ in comparison with $F_{\lambda_i}(\gamma_{L-1})$ and (b) is
obtained by combining (\ref{eq-last}) and (\ref{eq-fi}). Note that (\ref{eq-l1}) is only a function of $\gamma_{L-2}$, solving
for $\gamma_{L-2}$ yeilds
\begin{equation*}
\gamma_{L-2} = \frac{c_{L-2}}{\p^{[1+(n-i+1)(m-i+1)]}}
\end{equation*}
Repeating the procedure for all the bins sequentially toward the first bin yields the statement of the lemma. \QED

So far the outage as a function of $\gamma_0$ and $\gamma_0$ as a function of $\p$ is derived. In order to complete the
discussion it is needed to find the relation between $\lambda_i$ and the multiplexing gain. To do so, let revisit the expression
for mutual information with quantized feedback (\ref{eq-mut1})
\begin{equation}
I(S;Y|Q(H)) = \sum_{k=1}^m \log\left(1 + \frac{P(\lambda_i)}{M} \lambda_k\right).\label{eq-mut2}
\end{equation}
Assume that a diversity order of $r$ is desired. If the $j^{th}$ term, $j = \lceil r\rceil$, in (\ref{eq-mut2}) is given enough
power, then the achieved rate is larger than $j \log(1+P(\lambda_j)\lambda_j/M)$. Because, there are $j-1$ term preceding the
$j^{th}$ term all of which have larger channel coefficient $\lambda_k, k<j$. That is equivalent of having a multiplexing gain of
$j\ge r$. If we assume that the $j^{th}$ term is of form $(r/j) \log(1+P(\lambda_i)\lambda_i/M)$ and allocate power based on
this rate, then the multiplexing gain of $r$ is guaranteed.

Similarly if a multiplexing gain of $r$ is desired, and $\log(1+P(\lambda_j))\lambda_j)<(r/j)\log(1+\p)$, then with high
probability the transmission is in outage. Although we showed that outage only depends on the distribution of the $\lambda_j$
but $\lambda_j$ need not to be the channel parameter at the receiver that being quantized and fedback. For example, channel
parameter could be $\lambda_i$, which determines the feedback and power levels at the transmitter, but a multiplexing gain $j<i$
is desirable. So the outage is determined by the distribution of $\lambda_j$, while $\gamma_0$ (the cutoff threshold) is
determined by the average power and the distribution of $\lambda_i$.

The probability of outage can be obtained by replacing (\ref{eq-g0i}) into (\ref{eq-cor1}) and the diversity order of a MIMO
system with $L$ level power control can be derived
\begin{equation}
d = \left(1-\frac{r}{i}\right) (n-i+1)(m-i+1) G(m, n, i, L).\label{eq-cor2}
\end{equation}
Note that the channel parameter $\lambda_i$ is being fedback, and the outage in (\ref{eq-o1}) depends on the distribution of
$\lambda_i$. Consider that a multiplexing gain of $j-1\le \eta<j$, with $j\le i-1$ is desired. Then only the first $j$ terms
(instead of $i$ terms) in (\ref{eq-mut2}) is needed to achieve the rate. Hence, the outage is determined by the distribution of
$\lambda_j$, i.e.,
\begin{eqnarray}
\Pi(R(P_{av})) &=&
\Pr\{\lambda_j<\gamma_{0i}\}\nonumber\\
&=& \gamma_{0i}^{(n-j+1)(m-j+1)}\nonumber\\
&=& \left(\frac{c}{P_{av}^{(1-\alpha)G(m,n,i,L)}}\right)^{(n-j+1)(m-j+1)}, \label{eq-newout}
\end{eqnarray}
for some constant $c$. Note that outage threshold is denoted as $\gamma_{0i}$ to remind that it depends on the quantization
parameter $\lambda_i$ and prevent confusion with $\lambda_j$. \QED


\section{Extensions and Discussion}
\label{sec:discuss}

\subsection{Joint Power and Rate Control}

In addition to power control, the finite feedback can be used for rate control or a combination of rate and power control. In
this section we investigate the effect of power and rate control by constructing an example and analyzing its diversity and
multiplexing tradeoff curve.

Let $\gamma_{th}$ be a threshold on the channel state parameter, $\lambda_i$, such that a) $\gamma_{th}$ vanishes as $P_{av}$
increases, and b) for channel states $\lambda_i>\gamma_{th}$ we use a codebook, $\mathcal{C}_1$, with rate $R_1(\p) =
r_1\log(\alpha \p)$ and power $\alpha P_{av}$, where $\alpha$ is a fixed constant (for all $P_{av}$'s) between zero and one, and
c) for channel states $\lambda_i<\gamma_{th}$ we use a codebook, $\mathcal{C}_{2}$, with rate $R_2(\p) = r_2\log((1-\alpha)\p)$.

Since $\gamma_{th}$ decreases as $\p$ increase, asymptotically only codebook $\mathcal{C}_1$ is being used, {\em i.e}, the
multiplexing gain is provided by the $\mathcal{C}_1$. A power allocation that guarantees outage free transmission of
$\mathcal{C}_1$ fundamentally separates the outage event from the multiplexing achieving code. Using the rest of the
quantization bins on quantized power control for $\mathcal{C}_2$ minimize the outage. However, as it is discussed in
Section~\ref{ref:quantize} outage is inevitable due to singularity of channel inversion at $\lambda_i = 0$. Having $r_1<m$
guarantees an outage probability that decays with SNR with a slope greater than zero. The above intuition is summarized in the
next theorem.

\begin{thm}[Diversity and multiplexing with power and rate control] A multiple antenna system with power and rate control
described above has diversity order and multiplexing gain given by
\begin{eqnarray*}
&d = (1-\frac{r_1}{i})(n-i+1)(m-i+1) G(m,n,i,L-1)\\
&0\le r \le i . \label{eq-d}
\end{eqnarray*}
\label{theo-zehi}
\end{thm}
{\bf Proof:} As $\p$ increases, the $\gamma_{th}$ decreases. Therefore, the asymptotic rate is
\begin{eqnarray}
R(\p) &=& r_1\log(\alpha \p) \Pr\{\lambda_i>\gamma_{th}\} + r_1\log((1-\alpha) \p)\Pr\{\lambda_i<\gamma_{th}\}\nonumber\\
&=& r_1\log(\alpha \p).
\end{eqnarray}
Thus, the multiplexing gain of such a system depends only on the rate of $\mathcal{C}_1$. If we choose $\gamma_{th}$ such that
the power allocated to the $\mathcal{C}_1$, $\alpha P_{av}$, guarantees a zero outage transmission, then the outage only happens
in the transmission of codeword of $\mathcal{C}_2$. Therefore, the diversity order is the same as developed in the previous
section, with the difference that the multiplexing gain is decoupled from the diversity order. Hence, $\mathcal{C}_2$ can have a
fixed rate, zero multiplexing. Therefore, for $r=i$, $\mathcal{C}_1$ achieves the highest multiplexing gain, while
$\mathcal{C}_2$ provides a non-zero diversity order. Note that in order to guarantee zero outage during the transmission of
$\mathcal{C}_2$, the rate of the code can not exceed
\begin{equation*}
R(\p) = r \log(1+ \alpha\p\gamma_{th}).
\end{equation*}
and throughput is given by
\begin{equation}
T(\p) = r \log(1+ \alpha\p\gamma_{th})\Pr\{\lambda_i>\gamma_{th}\}.\label{eq-tt1}
\end{equation}

Although we do not aim to choose a threshold $\gamma_{th}$ that maximizes the throughput and we only care about the multiplexing
gain, an approximation to the threshold that maximizes the total throughput given in (\ref{eq-tt1}) could be a good candidate
for $\gamma_{th}$. Since $\Pr(\gamma>\gamma_{th})$ is monotonically decreasing with $\gamma_{th}$ and $\log(1+\alpha P_{av}
\gamma_{th})$ is monotonically increasing with $\gamma_{th}$, there is a $\gamma_{th}\in \mathcal{R}^+$ that maximizes
(\ref{eq-tt1}). Consider that $\gamma$ has the distribution of $\lambda_i$, for some $i$, $1\leq i\leq m$, and asymptotic
approximation given in Lemma~\ref{lem-cdf}, then by setting the first derivative of (\ref{eq-tt1}) with respect to $\gamma_{th}$
to zero we have
\begin{equation}
\frac{\alpha P_{av}}{1+\alpha P_{av}\gamma_{th}}-\log(1+\alpha P_{av} \gamma_{th}) f(\gamma_{th}) = 0, \label{eq-dr0}
\end{equation}
where $f(\cdot)$ is probability density function of $\lambda_i$. Replacing $f(\gamma_{th})$ by approximation of (\ref{eq-f})
around the origin we get
\begin{equation}
\gamma_{th} \approx \frac{1}{(\log(\alpha P_{av}))^{1/k}}, \label{eq-dr1}
\end{equation}
where $k=(n-i+1)(m-i+1)$. Note that
\begin{equation}
r_i = \lim_{\alpha P_{av}\rightarrow\infty}\frac{\log\left(1+ \frac{\alpha P_{av}}{(\log(\alpha P_{av}))^{1/k}}\right)}{\log
(\alpha P_{av})}= r, \label{eq-rtest}
\end{equation}
as expected.  Since for channel states greater than $\gamma_{th}$ the channel can support higher rates, it is guaranteed that
for this choice of threshold and power, the transmission of codewords from variable rate code is outage free. The diversity
order however, is the same as in (\ref{eq-dmax}) with consideration that one of the quantization thresholds is used for rate
control.\QED

\begin{figure}[hbtp]
\centering \epsfysize=5in \epsffile{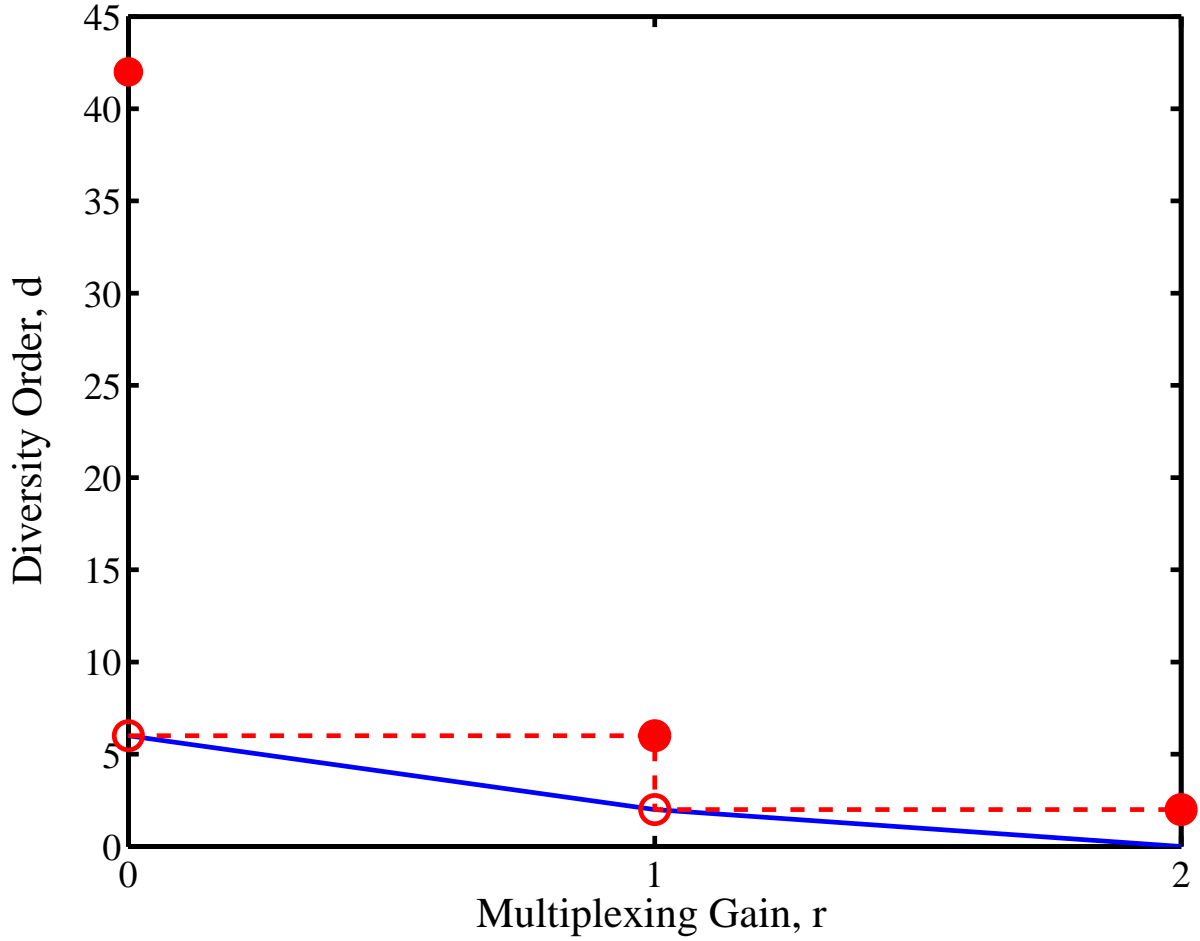} \caption{\small{Diversity-Multiplexing curve for systems with no and
partial CSI at the transmitter with M=2 transmit and N=3 receive antennas, and B=1 bit of feedback. Solid curve is corresponding
to the system with CSI only at the receiver. Dashed curve with circles corresponds to the system with perfect CSI at the
receiver and partial knowledge at the transmitter. The filled circles and empty circles are corresponding to the closed and open
end of the intervals respectively.}}\normalsize\vspace{-.1in}\label{fig-rd1}
\end{figure}

%
%
Figure~\ref{fig-rd1} shows multiplexing-diversity curve derived from the results of Theorem~\ref{theo-zehi} along with
multiplexing-diversity curve for a system with no CSI at the transmitter. It is assumed that $m=\min(M,N)=2, n=\max(M,N)=3$, and
there are $B=1$ bits of feedback (which corresponds to a quantizer with 2 bins). The three points on the (multiplexing,
diversity) curve for the system with one bit of feedback are $(0,42), (1,6)$ and $(2,2)$. Note that for a system with full
multiplexing with finite rate feedback a nonzero diversity order is achievable.

\subsection{Location-dependent Power Control}

Temporal power control is employed in many systems such that the power control commands,  which direct the transmitter to use a
certain power level, are \emph{independent} of average SNR of the channel. Average SNR is related to spatial location of the
receiver in relation to the transmitter and depend on the mobility of the mobile and/or the environment. This implies that the
power control thresholds are independent of receiver location, which is a judicious design for mobile systems.

However, in newer applications which are using MIMO links for point-to-point stationary links~\cite{KSK03}, the average
conditions of the links do not vary that rapidly. Thus, the two nodes in the system could adapt their power control commands
based on their relative location or equivalently average channel conditions. This leads to a system where  the feedback link
depends on the forward link average conditions, leading to much lower outage probabilities compared to a system which always
uses the same power control commands for all forward link conditions. We believe that control or side channel adaptation is
fundamentally required to improve on current systems which either have no side information or have a non-adaptive side
information.





\section{Conclusions}
\label{se:conclude}

In this paper we studied the effect of channel side information at the transmitter. There are four major messages of this paper.
First outcome indicates that power control (allocation of power in time) is much more effective then beamforming (spatial power
allocation). While the beamforming schemes has the same diversity order as systems without CSIT, even few bits of information at
the transmitter have a significant increase on diversity order and reliability. Second implicit result of the paper is that with
perfect channel state information with long term power control there is no tradeoff between multiplexing gain and diversity
order. Except for special cases, for almost all configurations of antennas, either probability of outage has an exponential
decay with SNR, or it is zero for finite SNR. Both cases have infinite diversity order. However, our third result of the paper,
shows that the tradeoff does exist when quantized channel state information is available at the transmitter. That is, the
diversity order with limited CSIT is finite, is a function of number of bits in feedback, and is much larger than diversity
order of system with CSIR only. However, the diversity order of a quantized power control system is zero at the maximum
multiplexing gain, and is independent of the feedback rate. Our fourth result implies that by applying a rate control mechanism
it is possible to achieve a non-zero diversity order at highest multiplexing gain.


\appendix
\section{Proof of Theorem~\ref{the-maxmux}}\label{app-p1}
Since the transmitter has no knowledge about the channel, it is inevitable to allocate same power $P=P_{av}/M$ on all the
transmit antennas. Therefore, the expression for mutual information can be written as
\begin{eqnarray}
I(X;Y) &=& \log\det\left(I_N + \frac{P_{av}}{M} HH^\dagger\right)\nonumber\\
&\stackrel{SVD}{=}& \log \det \left( I_N +  \frac{P_{av}}{M}  (U \Lambda^{1/2} V^\dagger )(U \Lambda^{1/2} V^\dagger)^\dagger\right),\nonumber\\
&=& \log \det \left( I_m + \frac{P_{av}}{M} \Lambda \right),\nonumber\\
&=& \log \left( \prod_{i=1}^m\left(1 + \frac{P_{av}}{M} \lambda_i\right) \right),\nonumber\\
&=&\log\left[\left(\frac{P_{av}}{M}\right)^m \left(\prod_{i=1}^m \lambda_i + \frac{M}{P_{av}}\sum_{i=1}^m\prod_{j\ne
i}\lambda_i+\cdots+\left(\frac{M}{P_{av}}\right)^m\right)\right]\label{eq-th2}
\end{eqnarray}
In the limit as $P_{av}\rightarrow\infty$, the expression for mutual information in (\ref{eq-th2}) is reduced to
\begin{equation*}
I(X;Y) = \log\left[\left(\frac{\p}{M}\right)^m\prod_{i=1}^m \lambda_i\right]
\end{equation*}
The probability of outage is
\begin{eqnarray*}
\Pi_{CSIR} &=& \Pr\{I(X;Y)<R\}\\
&=& \Pr\left\{\log\left[\left(\frac{\p}{M}\right)^m\prod_{i=1}^m \lambda_i\right]\le m\log\p\right\}\\
&=& \Pr\left\{\prod_{i=1}^m \lambda_i\le\frac{1}{M^m}\right\}.
\end{eqnarray*}
The outage probability is independent of $\p$ and is non-zero. Therefore, the total achievable rate can not exceed
$(1-\Pi_{CSIR})R = m(1-\Pi_{CSIR})\log\p$. Thus the multiplexing gain is at most $m(1-\Pi_{CSIR})$. \QED

\section{Proof of Theorem~\ref{the-csirt}}
\label{test} The optimum power control for a MIMO system with long term power control is found in~\cite{BCT01} and given by
\begin{equation}
P_i = \left[\left(\frac{2^{R/m}}{\left(\Pi_{j=1}^m\lambda_j\right)^{1/m}}\right)-\frac{1}{\lambda_i}\right]^+.\label{eq-optpc}
\end{equation}
There are some conditions regarding positiveness of $P_i$ which are not important in an asymptotic regime when $R$ can be assume
large enough that $P_i$'s are all positive for all $i\in\{1,2,\dots,m\}$. Having $R=r\log(P_{av})$, the outage region of the
optimum power control is defined as the channel realizations for which power is not allocated, i.e.
\begin{equation}
\Omega_{off} = \left\{\Lambda: \mathds{E}\left[\sum_{i=1}^m P_i\right]>\p\right\}.\nonumber
\end{equation}
From (\ref{eq-optpc}) we have
\begin{eqnarray}
P_1 &=&  \left(\frac{\p^{r/m}}{\left(\Pi_{j=1}^m\lambda_j\right)^{1/m}}\right)-\frac{1}{\lambda_1}\nonumber\\
&<& \frac{\p^{r/m}}{\lambda_1}-\frac{1}{\lambda_1}\nonumber\\
&\approx&\frac{\p^{r/m}}{\lambda_1} \;\;\;\;\;\;\text{for $\p$ sufficiently large,}\nonumber
\end{eqnarray}
which is an approximation to the power corresponding to the largest eigenvalue (assuming ordered eigenvalues with $\lambda_1$
the largest). We can obtain a similar approximation for $P_m$, the power corresponding to the smallest eigenvalue, i.e.,
\begin{equation}
P_m \approx \frac{\p^{r/m}}{\lambda_m}.\label{eq-pm}
\end{equation}
Using $P_1$ and $P_m$ define
\begin{equation}
\underline{B} = \left\{\Lambda: \mathds{E}\left\{m P_1\right\}>\p\right\}\text{      and      }\overline{B} = \left\{\Lambda:
\mathds{E}\left\{m P_{m}\right\}>\p\right\}.
\end{equation}
Then it is clear that
\begin{equation}
\underline{B}\subseteq\Omega_{off}\subseteq\overline{B}.
\end{equation}
Therefore, probability of outage can be bounded by
\begin{equation}
\Pr\left\{\underline{B}\right\}\le \Pi(\p, R(\p))\le\Pr\left\{\overline{B}\right\}. \label{eq-lulim}
\end{equation}
Let $\gamma_0^1$ be the solution to
\begin{equation}
\int_{\gamma_0^1}^\infty m P_1 f_{\gamma_1}(x) dx=\p,
\end{equation}
and $\gamma_0^m$ be the solution to
\begin{equation}
\int_{\gamma_0^m}^\infty m P_m f_{\gamma_m}(x) dx=\p,\label{eq-g0m}
\end{equation}
then inequalities in~(\ref{eq-lulim}) can be expanded as
\begin{equation}
\Pr\{\lambda_1<\gamma_0^1\}\le\Pi(\p,R(\p))\le\Pr\{\lambda_m<\gamma_0^m\}.\label{eq-drinq}
\end{equation}
\begin{itemize}
\item Case 1:($r<m$)\\
Replace the distribution of the smallest eigenvalue~\cite{Ede89} and~(\ref{eq-pm}) in (\ref{eq-g0m}) and solving for
$\gamma_0^m$ we have
\begin{eqnarray}
\p &=& \int_{\gamma_0^m}^\infty m \frac{\p^{r/m}}{x} \frac{x^{n-m} e^{-x}}{\Gamma(n-m+1)} dx\nonumber\\
\p^{1-r/m} &=& \frac{m}{\Gamma(n-m+1)}\int_{\gamma_0^m}^\infty x^{n-m-1} e^{-x} dx\nonumber\\
&<& \frac{m}{\Gamma(n-m+1)}\int_0^\infty x^{n-m-1} e^{-x} dx\nonumber\\
&=& \frac{m \Gamma(n-m)}{\Gamma(n-m+1)}\nonumber\\
&=& \frac{m}{n-m}.\label{eq-g0nm}
\end{eqnarray}
Therefore, for sufficiently large $\p$, to be specific for $\p>(m/(n-m))^{m/(r-m)}$, $\gamma_0^m$ is zero, which is equivalent
of saying $\overline{B}=\{\}$, which in turn implies that $\Pr\{\overline{B}\}=0$. From (\ref{eq-lulim}) and asymptotic
emptiness of $\overline{B}$ we conclude
\begin{eqnarray}
d &>&\lim_{\p\rightarrow\infty}-\frac{\log\left(\Pr\left\{\overline{B}\right\}\right)}{\log(\p)}\nonumber\\
&=& \infty.\label{eq-divinf}
\end{eqnarray}
The argument is not yet complete and requires more careful consideration for $n=m$. For the case of $n=m$ look at the single
antenna setting below.
\item Case 2:($r<m$ and $n=m$)\\
For a single antenna system or when number of transmit and receive antennas are equal, smallest (or the only) eigenvalue has the
exponential distribution, i.e., $f_{\lambda_m}(x) = e^{-x}$. Solving~(\ref{eq-g0m}) for $\gamma_0^m)$ we have
\begin{eqnarray}
\p^{1-r/m} &=& \int_{\gamma_0^m}^\infty m\frac{e^{-x}}{x} dx\nonumber\\
&\approx& m \log(x)\;\;\;\;\;\;\;\;\text{for small }\gamma_0^m.\label{eq-g01}
\end{eqnarray}
Therefore, $\gamma_0^m\approx e^{-\p^{1-r/m}/m}$. Now we can find $\Pr\{\overline{B}\}$ by
\begin{eqnarray}
\Pr\{\overline{B}\} &=& \Pr\{\lambda_m<\gamma_0^m\}\nonumber\\
&=& \int_0^{\gamma_0^m} e^{-x} dx\nonumber\\
&=& e^{-\gamma_0^m}\nonumber\\
&=& 1 - e^{- e^{-\frac{\p^{1-r/m}}{m}}}\nonumber\\
&\approx& e^{-\frac{\p^{1-r/m}}{m}}.\nonumber
\end{eqnarray}
Which yields same diversity order as in~(\ref{eq-divinf})
\item Case 3:($r=m$)\\
From (\ref{eq-g01}) and (\ref{eq-g0nm}) it is immediate that $\gamma_0^m = c_m$, where $c_m$ is a constant. Thus,
$\Pr\{\overline{B}\}$ is constant and independent of $\p$. If $\Pr\{\overline{B}\}$ is zero, then there is no outage, and a
diversity order of infinity is achieved, but if $\Pr\{\overline{B}\}$ is not zero, then we need to find the
$\Pr\{\underline{B}\}$.

Case of $n=m=1$ which is the single antenna case is treated in the previous case and it is shown that the probability of outage
is constant, and hence the upper bound for the diversity order for a single antenna system is also zero. Thus the maximum
diversity order for a single antenna system at maximum multiplexing gain $r=m$ is zero.
\item Case 4:($r=m, n>1, m=1$)
This case includes MISO and SIMO systems. The channel is defined by a single scalar parameter which is the magnitude of the
channel vector with gamma distribution. The boundary of the outage region can be found from the average power constraint, i.e.,
\begin{eqnarray}
\p &=& \int_{\gamma_0}^\infty \frac{\p}{x} \frac{x^{n-1} e^{-x}}{\gamma(n)} dx\nonumber\\
1 &=& \int_{\gamma_0}^\infty \frac{x^{n-2} e^{-x}}{\gamma(n)} dx.\label{eq-g0miso}
\end{eqnarray}
For $\gamma_0 = 0$ Equation~(\ref{eq-g0miso}) is reduced to $1=1/(n-1)$, and the outage region is defined by all channel states
$\lambda$ for which $1/(n-1)>1$, which is an empty set. Therefore, the outage is zero and diversity order of infinity can be
achieved at the maximum multiplexing gain of 1.
\item Case 5:($r=m, n,m >1$)\\
By having $r=m$ in (\ref{eq-g0nm}) we get that for $n>2m$, $\overline{B}$ is empty, and so is probability of outage, and the
diversity order is infinite. However, for $m\le n\le 2m$, on one hand $\overline{B}$ is not empty, and on the other hand the
distribution of $\lambda_1$ is not known. So it is not clear that whether the probability of outage is zero or non-zero. We need
to see if $\Omega_{off}$ is empty, and if it depends on $\p$.

For large values of $\p$, we can write the power corresponding to each eigenvalue in (\ref{eq-optpc}) as
\begin{equation}
P_i = \frac{\p}{(\Pi_{j=1}^m \lambda_j)^{1/m}}.
\end{equation}
For large $\p$ all powers all almost equal and the term $1/\lambda_i$ is negligible. The outage region, $\Omega_{off}$ is
determined by the solution of $\mathds{E}[m P_i]=\p$, for some $i$, i.e.,
\begin{equation}
\mathds{E}\left[\frac{1}{(\Pi_{j=1}^m \lambda_j)^{1/m}}\right] = \frac{1}{m}\label{eq-fff},
\end{equation}
which shows that the outage region does not depend on $\p$. Therefore, probability of outage is constant. If the outage region
is empty, then probability of outage is zero and diversity order is infinite, and if outage region is non-empty, then diversity
order is zero. Using Jensen's inequality we can further simplify (\ref{eq-fff}) to get
\begin{eqnarray}
\mathds{E}\left[\frac{1}{(\Pi_{j=1}^m \lambda_j)^{1/m}}\right]&\ge&\mathds{E}\left[(\Pi_{j=1}^m \lambda_j)^{1/m}\right]\nonumber\\
&=& \mathds{E}[\lambda_j],
\end{eqnarray}
for an arbitrary unordered eigenvalue $\lambda_j$.
\end{itemize}
\QED

\bibliographystyle{IEEEtran}
\bibliography{bib/tran}

\newpage


\end{document}